\documentclass[11pt]{article}
\usepackage[utf8]{inputenc}
\usepackage[T1]{fontenc}
\usepackage{graphicx}
\usepackage{float}
\usepackage{amsmath}
\usepackage{amssymb}
\usepackage{hyperref}
\usepackage{subcaption}
\usepackage{framed}
\usepackage{tabularx}
\usepackage{mathtools}
\usepackage{tikz}
\usetikzlibrary{decorations.pathreplacing} 
\usetikzlibrary{patterns}

\DeclareMathOperator*{\argmax}{arg\,max}

\author{Stefano Bennati, Leonel Aguilar, Dirk Helbing}
\date{}
\title{How intelligence can change the course of evolution.}
\begin{document}

\maketitle

\begin{abstract}
  The effect of phenotypic plasticity on evolution, the so-called Baldwin effect, has been studied extensively for more than 100 years.
  Plasticity is known to influence the speed of evolution towards a specific genetic configuration, but whether it also influences what that genetic configuration is, is still an open question.
  This question is investigated, in an environment where the distribution of resources follows seasonal cycles, both analytically and experimentally by means of an agent-based model of a foraging task.
  Individuals can either specialize to foraging only one specific resource type or generalize to foraging all resource types at a low success rate.
  It is found that the introduction of learning, one instance of phenotypic plasticity, changes what genetic configuration evolves.
Specifically, the genome of learning agents evolves a predisposition to adapt quickly to changes in the resource distribution, under the same conditions for which non-learners would evolve a predisposition to maximize the foraging efficiency for a specific resource type.
  This paper expands the literature at the interface between Biology and Machine Learning by identifying the Baldwin effects in cyclically-changing environments and demonstrating that learning can change the outcome of evolution.
  \end{abstract}

\section{Introduction}

The so called Baldwin effect \cite{baldwin96_new_factor_evolut} is a much debated theory in the literature of evolution \cite{ancel00_under_baldw_exped_effec} about how new features are inherited by an individual with phenotypic plasticity \cite{west1989phenotypic,dewitt2004phenotypic,via1995adaptive}.
Baldwin proposed this new ``factor in evolution'' \cite{baldwin96_new_factor_evolut} to explain how complex features such as an eye can evolve \cite{sterelny04_review_evolut_learn,dejager16_baldw_remar_effec,crispo2007baldwin}, as an alternative to the then-popular Lamarckian evolution which assumed that traits acquired by an individual through phenotipic plasticity would be transferred directly to its offspring's genome \cite{burkhardt2013lamarck}.
This idea got unnoticed until the late 1990s, when it caught the interest of the fields of Psychology, in reference to the evolution of human learning, and Computer Science, in reference to evolutionary computation and machine learning. Only from the mid-2000s the Baldwin effect started taking ground in the field of Evolutionary Biology. \cite{scheiner2014baldwin}.

Learning, i.e. an instance of phenotypic plasticity \cite{anderson1995learning,dennett03_baldw_effec,whitman2009phenotypic}, has been found to affect how evolution reaches an optimal configuration \cite{french1994genes,hinton87_how_learn_can_guide_evolut} by either speeding up \cite{nolfi99_learn_evolut} or slowing down the evolutionary process \cite{ancel00_under_baldw_exped_effec,paenke09_influen_learn_evolut}.
This work brings that concept a step further by demonstrating that learning can change the outcome of the evolutionary process.
The effect of learning on evolution is studied both experimentally, by means of an Agent-Based model of a foraging task  \cite{hamblin09_findin_evolut_stabl_learn_rule,redko10_learn_evolut_auton_adapt_agent}, and analytically, by means of a mathematical model \cite{sznajder11_how_adapt_learn_affec_evolut}.
Foraging of different resource types is subject to trade-offs: the more an agent specializes in one resource, the less effectively it can forage the other resource, e.g. due to neophobia \cite{beissinger94_exper_analy_diet_special_snail_kite}, a non-transferable skill set or other constraints, e.g. energy or memory constraints.
This trade-off is modeled by a single parameter that determines the probability of success of foraging two resource types \cite{laverty1994costs}.
Specifically, \emph{aptitude} defines the parameter value encoded in the genome and inherited by the parent, while \emph{skill} defines the corresponding phenotypic expression which determines the probability of successful foraging.
Aptitude changes from one generation to the next due to random mutations, and learning allows the skill to change from the inherited aptitude to a value more suited to the current state of the environment.
The environment cycles periodically between two different configurations, named \emph{seasons} \cite{pulliam92_popul_dynam_compl_lands,dridi15_envir_compl_favor_evolut_learn,hamblin09_findin_evolut_stabl_learn_rule}, which determine what resources are available for agents to forage.
The choice of model favored simplicity over realism, modeling realistic entities and ecosystems is outside the scope of this work.

Computational experiments demonstrate the existence of the Baldwin effect in a cyclical environment, which can lead to both a speed up \cite{hinton87_how_learn_can_guide_evolut,fontanari1990effect} and a slow down \cite{ancel00_under_baldw_exped_effec} of the evolutionary process.
Further experimental and analytical results demonstrate that learning is not only able to condition the speed of convergence but also the evolved genetic configuration; we name this effect the \emph{Baldwin veering effect}.
Specifically it is found that, under the same conditions where agents with a fixed phenotype would evolve a specialist configuration, learning agents evolve a generalist configuration.
\emph{Specialist configuration} is defined as a genome whose aptitude evolves to one of the extreme values, while \emph{generalist configuration} is defined as a genome whose aptitude evolves to an intermediate value.
Analytical results confirm that learning changes the fitness landscape in a way that makes a generalist configuration a global optimum in the space of genotypes.
The intuition is that non-learning agents cannot adapt to the fast changes in the environment, so they maximize their foraging efficiency for one type of resource.
Conversely, learning agents can adapt to any environmental condition, and a generalist strategy offers them higher flexibility.

  The main contributions of this paper are to show that in a cyclically changing environment: (I) the well-known Baldwin effect is present, (II) the presence of learning affects the outcome of the evolutionary process by driving evolution to a different configuration, we name this the \emph{Baldwin veering effect}, (III) a mathematical model captures this new effect and confirms the experimental findings, and (IV) the existence of this new effect is conditioned only upon the relation between the speed of learning and the frequency of change in the environment.

This paper is structured as follows: Section \ref{lab:results} presents the design and results of the experiments substantiating the claims in this paper, Section \ref{lab:simple_model} introduces the analytical model and describes results that validate the experimental findings, Section \ref{lab:discussion} presents a short discussion, Section \ref{lab:conc} provides concluding remarks to this work and Section \ref{lab:method} concludes the paper by presenting details of the agent based methodology, the environmental setting and the learning mechanisms used for the computational experiments.

\section{Results of the computational experiments \label{lab:results}}

Previous work in the literature about the Baldwin effect found that learning can either speed up or slow down the evolutionary process, depending on the learning mechanism, the fitness function and the starting conditions of the population \cite{ancel00_under_baldw_exped_effec}.
The goal of this experiment is to verify whether or not the Baldwin effect exists in a cyclical environment, a question that, to the best of our knowledge, has not been answered before \cite{sznajder11_how_adapt_learn_affec_evolut}.

The existence of the Baldwin effect is evaluated by comparing the speed of genetic assimilation of phenotypic features that change in adaptation to changes in the environment.
Three populations are compared:
\begin{itemize}
\item reactive agents, i.e. unable to learn, is taken as baseline.
\item agents that can modify their own actions through learning (speed up).
\item agents which can modify their own actions and their skill through learning (slow down).
\end{itemize}

\begin{figure}[H]
  \centering
  \includegraphics[width=0.9\textwidth]{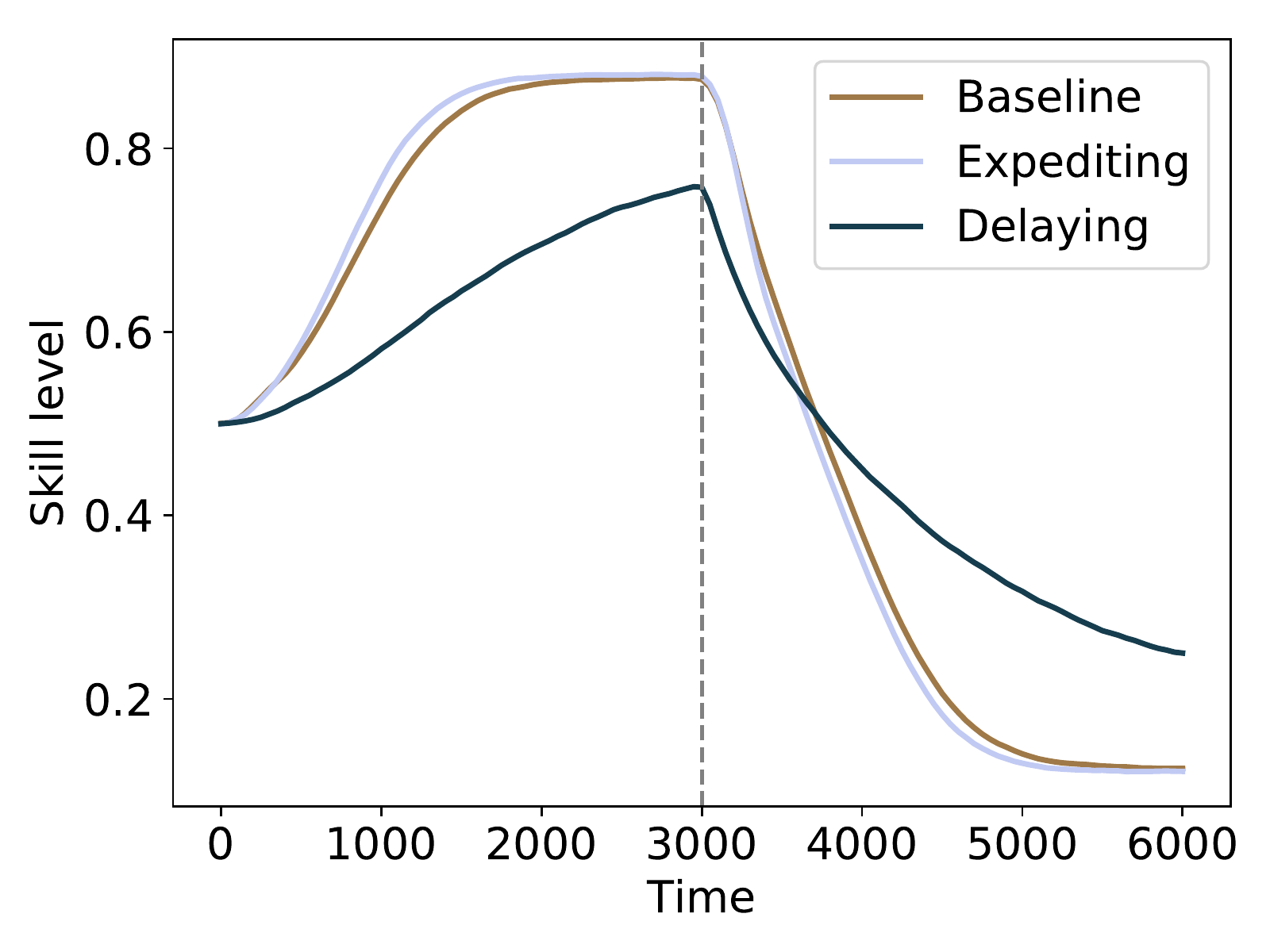}
  \caption{{\bf The Baldwin effect.} Evolution of aptitude over time in two different learning populations, compared to a baseline of reactive agents. The speed of genetic assimilation changes with respect to the baseline, depending on the configuration of the learning algorithm, demonstrating that the Baldwin can speed up or slow down the genetic assimilation of aptitude. The dashed vertical line indicates a change of season, i.e. resource availability. Confidence intervals at the 95\% confidence level are not shown as their size is negligible.}
  \label{fig1}
\end{figure}

Figure \ref{fig1} shows that the Baldwin effect is present as the speed of assimilation is affected by learning.

Learning allows agents to improve their foraging capacity over time by learning the correct mapping between actions and perceptions.
If the skill cannot be learned, individuals are selected based on their inherited aptitude value.
If the skill can be learned, the aptitude value determines only indirectly the individual's fitness which slows down its genetic assimilation.

\subsection{A new effect: the Baldwin veering effect}

This experiment investigates whether the Baldwin veering effect exists: the genetic configuration evolved in a cyclically-changing environment is different in presence or absence of reversible plasticity i.e. learning.
Unlike previous work, the frequency of change in the environment plays a crucial role \cite{Nolfi1996}, while the co-evolution of  different populations does not \cite{fordyce2006evolutionary,chakra2014plastic}.

The intuition is that learning helps natural selection traversing the space of genetic configurations, and does so at faster timescale.
Natural selection is able to adapt to slowly-changing environments, in this case learning might speed up or delay this process.
If instead the environment changes faster than the evolutionary timescale, learning and natural selection do not merely tend to the same objective but instead take on two different roles: Learning optimizes the behavior of agents in response to environmental variability, while natural selection optimizes the efficiency of learning.
Different genetic configurations correspond to different initial learning efforts; given that an individual has the same probability of being born in either season, the optimal genetic configuration should equally reduce the effort or learning either skill.

This prediction is verified by comparing the inherited aptitude across two different populations, one of learning agents and one of reactive agents.
The Baldwin veering effect is present if the two populations evolve a different genetic configuration, namely the reactive population specializes in either resource types while the learning population evolves a generalist configuration.

\begin{figure}[H]
  \centering
  \includegraphics[width=\textwidth]{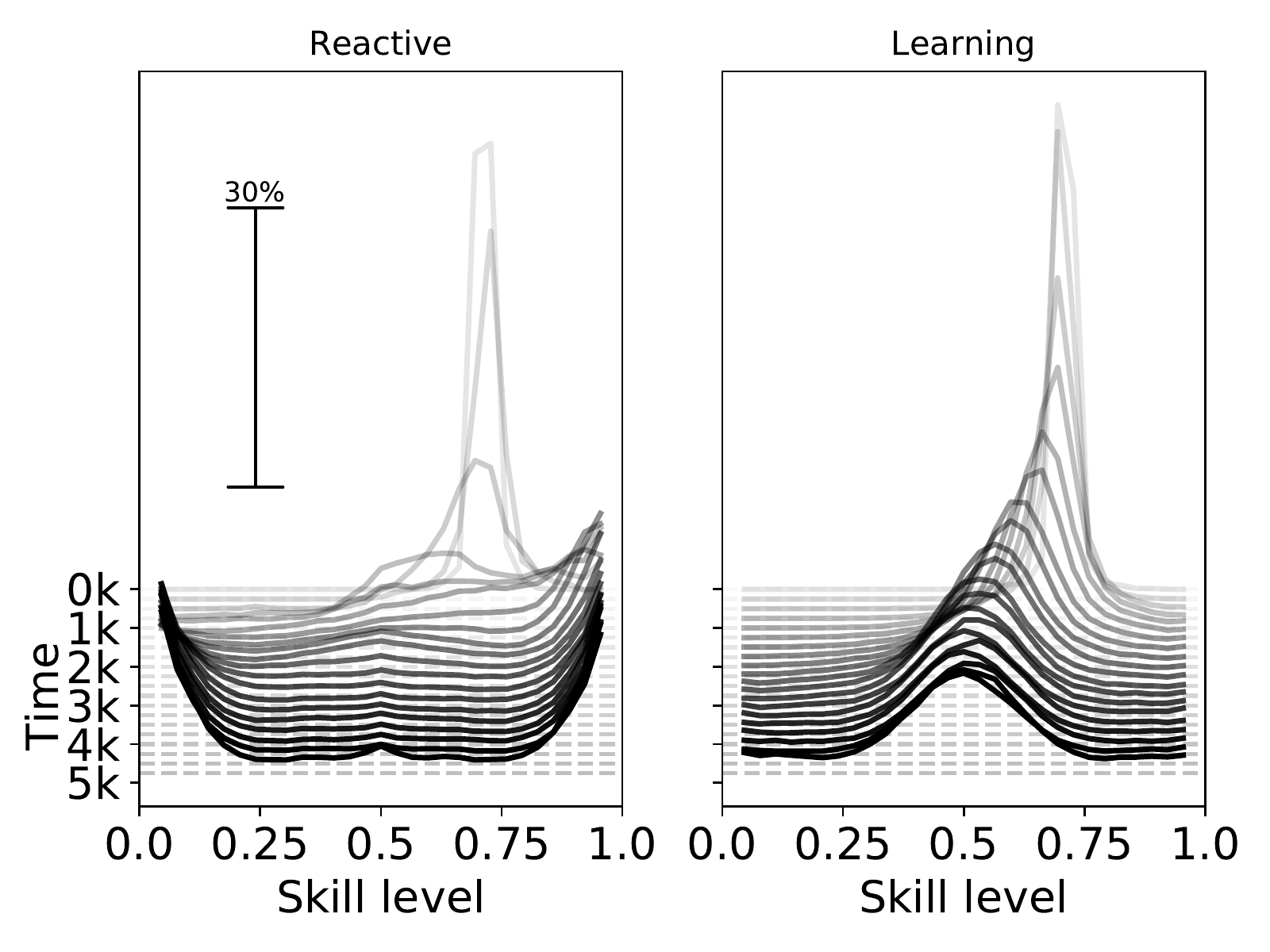}
  \caption{{\bf Comparison of aptitude distributions across experiments.} The graphs show the change in the distribution of aptitude values (x-axis) over the course of the simulation (z-axis). Each point in the graph represents the frequency at which a specific aptitude value was present in the population (y-axis) at a given time.
    The left plot shows a population of reactive agents while the right plot a population of plastic agents, the populations evolve two different distributions, confirming that learning can change the outcome of evolution. \label{fig2}}
\end{figure}

\begin{figure}[H]
  \centering
  \includegraphics[width=\textwidth]{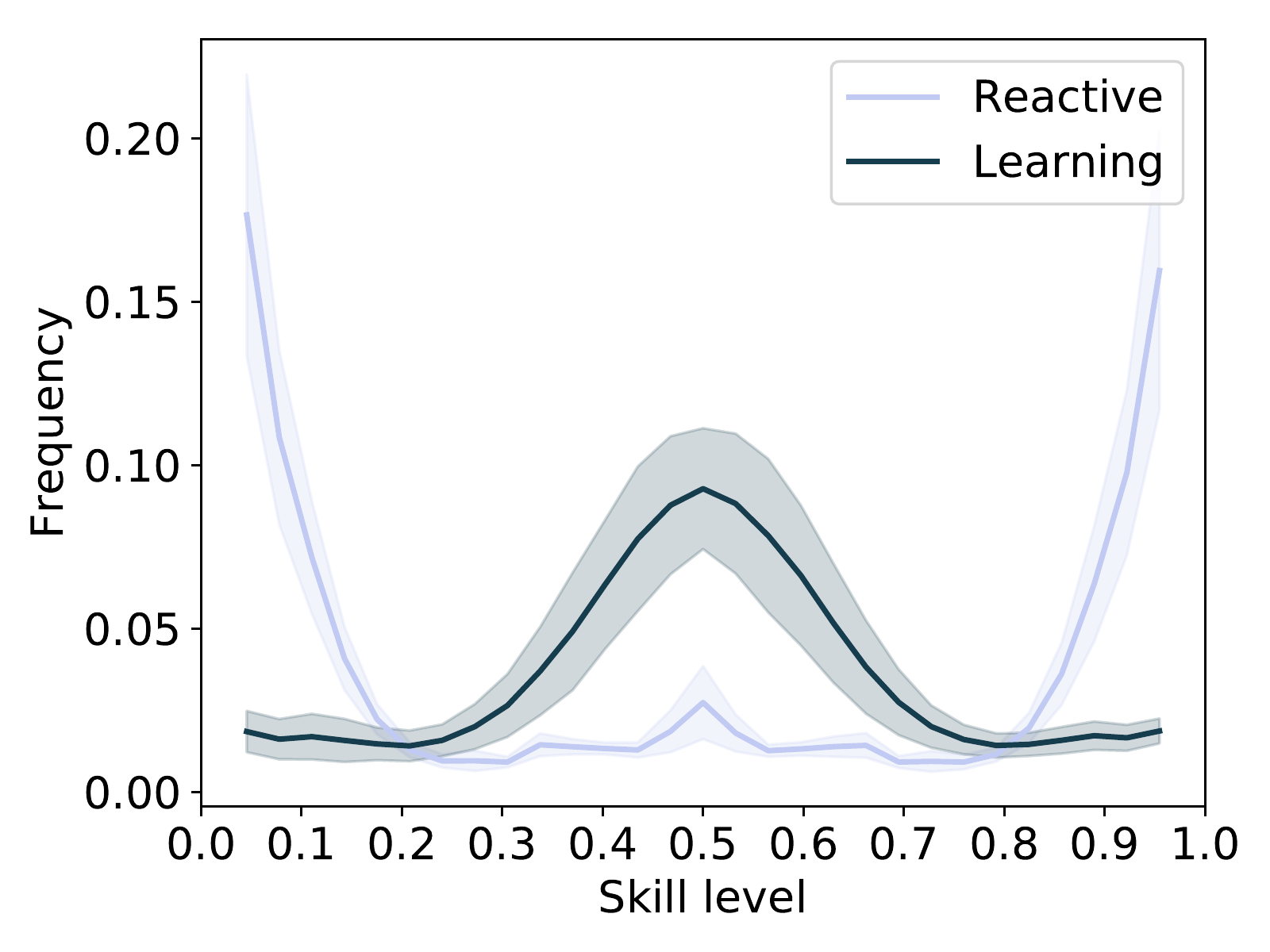}
  \caption{{\bf Comparison of final aptitude distribution across experiments.} The graphs show the average distribution of aptitude values (x-axis) for the last 1000 iterations of the simulation, shaded areas represent confidence intervals at the 95\% confidence level. The lines indicate the frequency at which a given aptitude occurs in the population. Plastic agents evolve a different distribution than reactive agents, confirming that learning can change the configuration to which evolution converges.\label{fig3}}
\end{figure}

Figure \ref{fig2} shows that a population of reactive agents evolves extreme aptitude values, i.e. a specialist configuration, thus each half of the population specializes in foraging one type of resource. A learning population instead evolves an intermediate aptitude value, i.e. a generalist configuration, which allows to adapt quickly to any environmental condition.
Figure \ref{fig3} highlights the difference between genetic configurations evolved by the two populations.

\subsection{Differences in individual behaviors}

In order to verify that a difference in genetic configuration actually results in different behaviors, the agents that are alive during the last timestep of the simulation are cloned and used to initialize a new set of simulations.
In these new simulations, the environment is set to have only one season and contains an equivalent quantity of both types of resources. Furthermore, agents do not reproduce and their behavior is fixed and fully determined by their genome.
In these new experiments the behavior of individuals is compared with the measures of foraging history and of group behavior, which are described in Section \ref{sec:measures}.

\begin{figure}[H]
\centering
  \includegraphics[width=\textwidth]{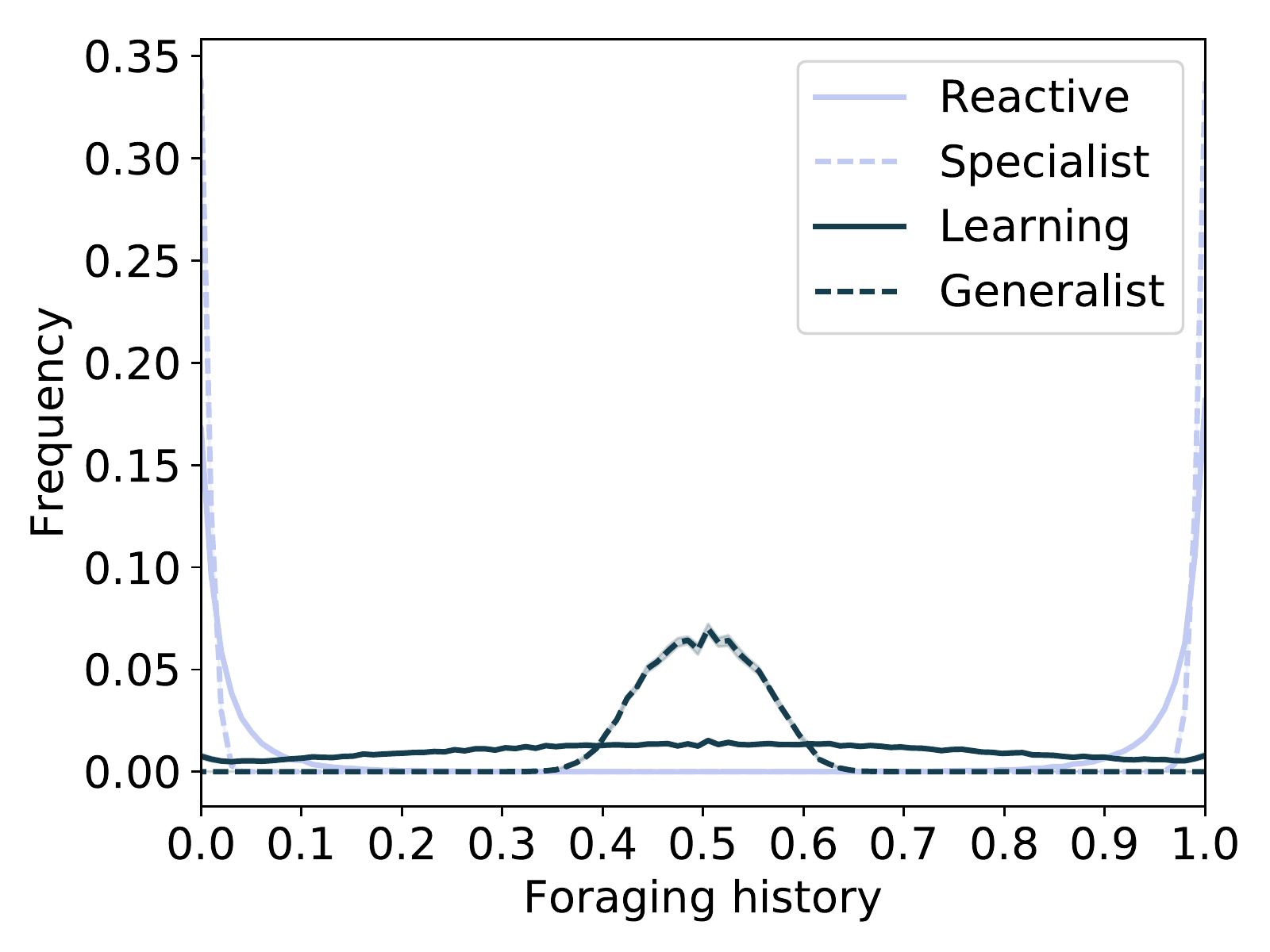}
  \caption{{\bf Comparison of foraging history.} The plots show the frequency at which a given value of foraging history occurs in the population. Foraging history is computed as the percentage of successful foraging actions or resources of type 0. A frequency of 0.2 associated to a value of foraging history of 0.4 means that 20\% of individuals in the population foraged during their lifetime 40\% of the time resources of type 0 and 60\% of the time resources of type 1.
  Distributions of foraging actions resemble the distributions of aptitudes, confirming that a difference in aptitude distribution corresponds to an actual difference in behavior. Dashed lines represent baseline populations, where all agents have aptitude of 0.5 (generalist configuration) or half of the population has aptitude 0.05 and the other half 0.95 (specialist configuration). Shaded areas (of negligible size) represent confidence intervals at the 95\% confidence level.\label{fig4}}
\end{figure}

The measure of foraging history shows that the behaviors in the two conditions differ (cf. Fig. \ref{fig4}), namely the reactive population splits in two groups of comparable size, each of which specializes in foraging one type of resource, while the learning population has a more uniform foraging pattern which includes more generalists.
The measure of individual foraging history is quantified by the frequency of foraging resources of type one, e.g. a value of 90\% indicates that 90\% of all resources foraged by the agent were of type one, and the remaining 10\% of type two.
These values are then aggregated across the population to determine the frequency of different values of foraging history.

\begin{figure}[H]
\centering
  \includegraphics[width=0.9\textwidth]{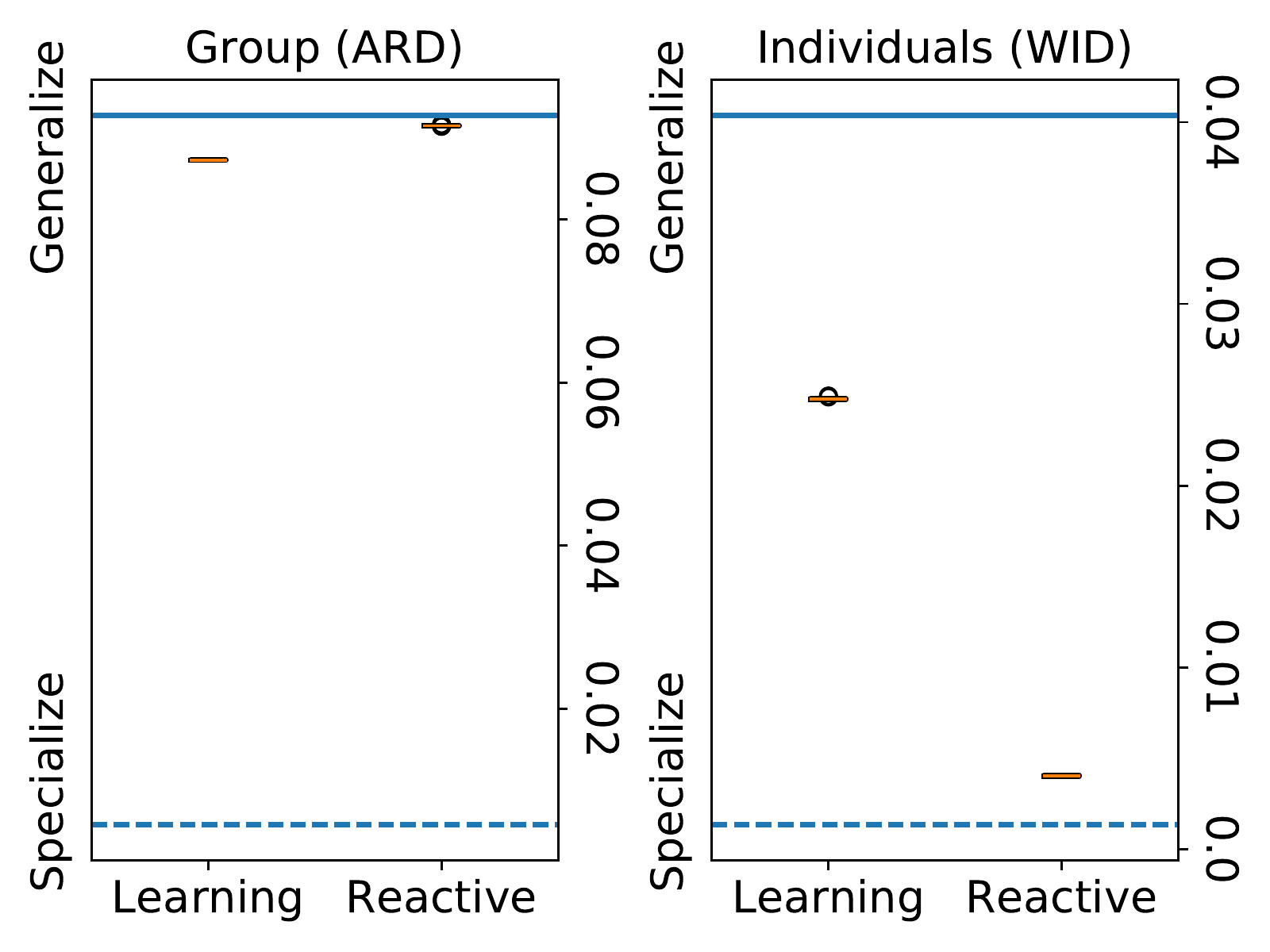}
\caption{{\bf Measures quantifying the behavior of the population.} Left: Among Resource Diversity quantifies the behavior of the population, both populations display a similar generalist behavior. Right: Within Individual Diversity quantifies the behavior or individual agents, learning agents behave more generalist than reactive agents. The solid line represents a baseline population in which all agents have skill of 0.5, the dashed line represents a baseline population in which each half of the agents has skill of 0.05 and 0.95 respectively. These results confirm that a difference in aptitude distribution corresponds to an actual difference in behavior. \label{fig5}}
\end{figure}

Besides the measure of foraging history, different standard measures of group behavior \cite[Pag.~241]{giraldeau00_social} are used to compare the behavior of the populations (cf Fig. \ref{fig5}).
The interpretation of these measures in not straightforward, so baselines are added for reference: the dashed line represents the value of a population where half of the agents specialize in one resource and the other half in the other resource, while the continuous line represents a population of generalists.

The measures confirm the results: learning agents develop a generalist foraging strategy, both at the group level (Among-Resource Diversity) and at the individual level (Within-Individual Diversity), while reactive agents develop a more specialized foraging strategy.
The result is not so clear for the group level of a reactive population, but it can be explained by including the result at the individual level:
The measure of Among Resource Diversity (ARD) is high either if different agents have different specialized diets or if agents generalize, and the result of Within Individual Diversity (WID) exclude the latter cause.

\section{A mathematical model \label{lab:simple_model}}

The results outlined in the previous section showcase the existence of the Baldwin veering effect, but give little information about the process behind it.
This section introduces and analyzes the predictions of an analytical model, inspired on previous work \cite{frankenhuis2016mathematical}, which give a possible explanation to the simulation results and identify the conditions under which the Baldwin veering effect manifests.
The model captures the individual fitness of agents through the definition of a general fitness function, the evolutionary process is not explicitly modeled so evolutionary outcomes are inferred from considerations about the relative fitness of different individuals.
More fine-grained results about evolution and its dynamics might be obtained by pairing the fitness function with any existing model of evolution, e.g. \cite{van1991evolution,frankenhuis2016mathematical}, such effort is outside the scope of this paper and is left for future work.

The environment contains two types of resources, whose proportion is denoted by $\alpha_0$ and $\alpha_1$.

The fitness of a reactive agent $i$ is formulated as follows:
\[
  W_i=\alpha_0 \cdot r_{i,0} + \alpha_1 \cdot r_{i,1}=\alpha_0 \cdot a_{i,0}^q+\alpha_1 \cdot a_{i,1}^q
\]

Where the foraging success $r_{i,j}=a_{i,j}^q$ is determined by the agent's aptitude $a_{i,j} \in [0,1]$ (which is equal to the skill level, being it a reactive agent) and by a parameter $q \in \mathbb N_{\ge 0}$ which defines the relation between skill and foraging success.
If the parameter $q=1$, specializing on one resource and generalizing on two resources lead to the same foraging success. If $q>1$ specialization is more beneficial as intermediate aptitude produce a lower foraging success than extreme ones, vice versa if $q<1$ generalization becomes more beneficial than specialization.

Following the design of the computational model, the two skills of an agent are assumed to be complementary, $a_{i,0}+a_{i,1}=1$ as well as the resource availability $\alpha_1+\alpha_0=1$.
Therefore, the notation can be simplified by defining $a_i \coloneqq a_{i,0}$ and $1-a_i \coloneqq a_{i,1}$.

\begin{equation}
W_i=\alpha_0 \cdot a_i^q+(1-\alpha_0) \cdot(1 - a_i)^q
\end{equation}

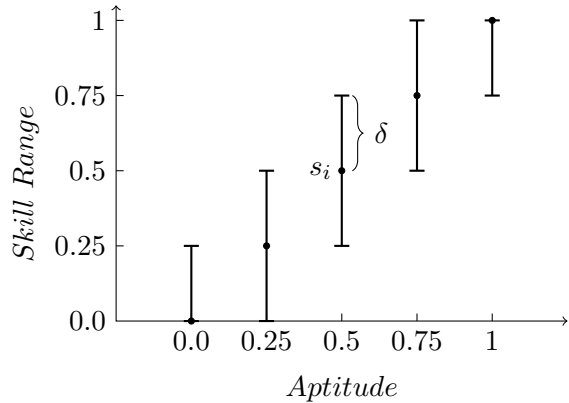
\begin{figure}[H]
  \centering
\begin{tikzpicture}[scale=2]
  \draw [<->] (0,2.1) -- (0,0) -- (3,0);
  \node [left,rotate=90] at (-0.6,1.5) {$Skill ~ Range$};
  \node [below] at (1.5,-0.3) {$Aptitude$};
  \node [left] at (0,0) {$0.0$};
  \draw [] (0,0.5) -- (0.05,0.5);
  \node [left] at (0,0.5) {$0.25$};
  \draw [] (0,1) -- (0.05,1);
  \node [left] at (0,1) {$0.5$};
  \draw [] (0,1.5) -- (0.05,1.5);
  \node [left] at (0,1.5) {$0.75$};
  \draw [] (0,2) -- (0.05,2);
  \node [left] at (0,2) {$1$};
  \draw [fill,color=black] (0.5,0) circle [radius=.02];
  \draw [thick] (0.45,0) -- (0.55,0);
  \draw [thick] (0.5,0) -- (0.5,0.5);
  \draw [thick] (0.45,0.5) -- (0.55,0.5);
  \node [below] at (0.5,0) {$0.0$};
  \draw [] (0.5,0) -- (0.5,0.05);
  \draw [fill,color=black] (1,0.5) circle [radius=.02];
  \draw [thick] (0.95,0) -- (1.05,0);
  \draw [thick] (1,0) -- (1,1);
  \draw [thick] (0.95,1) -- (1.05,1);
  \node [below] at (1,0) {$0.25$};
  \draw [] (1,0) -- (1,0.05);
  \draw [fill,color=black] (1.5,1) circle [radius=.02];
  \draw [thick] (1.45,0.5) -- (1.55,0.5);
  \draw [thick] (1.5,0.5) -- (1.5,1.5);
  \draw [thick] (1.45,1.5) -- (1.55,1.5);
  \node [below] at (1.5,0) {$0.5$};
  \draw [] (1.5,0) -- (1.5,0.05);
  \node [left] at (1.5,1) {$s_i$};
  \draw [decorate,decoration={brace,amplitude=4pt,mirror,raise=4pt},yshift=0pt]
(1.5,1) -- (1.5,1.5)node [black,midway,xshift=15pt] {$\delta$};
  \draw [fill,color=black] (2,1.5) circle [radius=.02];
  \draw [thick] (1.95,1) -- (2.05,1);
  \draw [thick] (2,1) -- (2,2);
  \draw [thick] (1.95,2) -- (2.05,2);
  \node [below] at (2,0) {$0.75$};
  \draw [] (2,0) -- (2,0.05);
  \draw [fill,color=black] (2.5,2) circle [radius=.02];
  \draw [thick] (2.45,1.5) -- (2.55,1.5);
  \draw [thick] (2.5,1.5) -- (2.5,2);
  \draw [thick] (2.45,2) -- (2.55,2);
  \node [below] at (2.5,0) {$1$};
  \draw [] (2.5,0) -- (2.5,0.05);
\end{tikzpicture}
  \caption{{\bf An explanation of the skill range $\delta$.} Skill ranges obtained with a fixed value of $\delta$ and different aptitudes.}
  \label{fig6}
\end{figure}

In order to model the effect of learning agents, a new parameter $\delta$ is introduced which represent plasticity.
A learning agent is not constrained by its inherited aptitude, which can be adapted to the conditions of the environment.
The value of $\delta$ determines the range of skills an agent can express, this range is centered in the aptitude and spans in both direction (cf. Fig. \ref{fig6}).

\begin{equation}
W_i=\alpha_0 \cdot min(1,(a_i+\delta))^q+(1-\alpha_0) \cdot min(1,(1 - a_i+\delta))^q -c \cdot \delta
\end{equation}

The parameter $c$ determines the cost of plasticity \cite{dewitt1998costs}. The skill cannot extend beyond the domain $[0,1]$, hence the bounding to $1$.

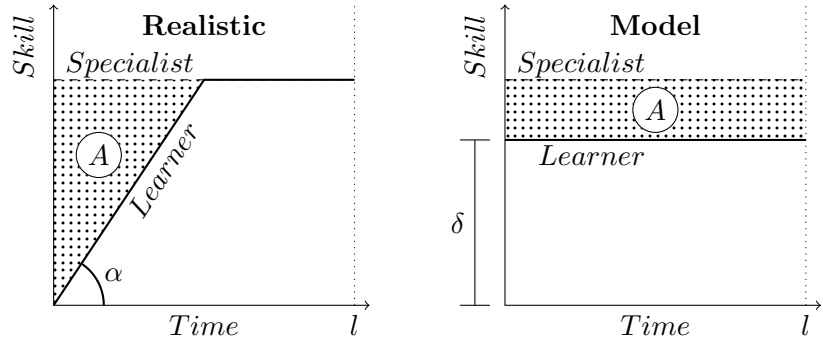
\begin{figure}[H]
  \centering
\begin{tikzpicture}[scale=2]
  \draw [<->] (0,2) -- (0,0) -- (2.1,0);
  \node [left,rotate=90] at (-0.2,2) {$Skill$};
  \node [below] at (1,0) {$Time$};
  \node [below] at (2,0) {$l$};
  \draw [dotted] (2,0) -- (2,2);
  \node [below,font=\bf] at (1,2) {Realistic};
  \draw [dashed] (0,1.5) -- (2,1.5);
  \node [left] at (1,1.6) {$Specialist$};
  \draw [thick] (0,0) -- (1,1.5) -- (2,1.5);
  \node [right, rotate=57] at (0.5,0.55) {$Learner$};
  \fill[pattern=dots] (0,0) -- (1,1.5) -- (0,1.5);
  \draw [fill,color=white] (0.3,1) circle [radius=.15];
  \draw [color=black] (0.3,1) circle [radius=.15];
  \node [] at (0.3,1) {$A$};
  \draw [thick,domain=0:55] plot ({cos(\x)/3}, {sin(\x)/3});
  \node [] at (0.4,0.2) {$\alpha$};
  \draw [<->] (3,2) -- (3,0) -- (5.1,0);
  \node [left,rotate=90] at (2.8,2) {$Skill$};
  \node [below] at (4,0) {$Time$};
  \node [below] at (5,0) {$l$};
  \draw [dotted] (5,0) -- (5,2);
  \draw [] (2.8,0) -- (2.8,1.1);
  \draw [] (2.7,0) -- (2.9,0);
  \draw [] (2.7,1.1) -- (2.9,1.1);
  \node [left] at (2.8,0.55) {$\delta$};
  \node [below,font=\bf] at (4,2) {Model};
  \draw [dashed] (3,1.5) -- (5,1.5);
  \node [left] at (4,1.6) {$Specialist$};
  \draw [thick] (3,1.1) -- (5,1.1);
  \node [left] at (4,1) {$Learner$};
  \fill[pattern=dots] (3,1.5) rectangle (5,1.1);
  \draw [fill,color=white] (4,1.3) circle [radius=.15];
  \draw [color=black] (4,1.3) circle [radius=.15];
  \node [] at (4,1.3) {$A$};
\end{tikzpicture}
  \caption{{\bf A sketch of modeling assumptions.} The graph shows the change of skill level over time of an hypothetical learning individual. The shaded area represents the cost of adaptation: the loss in fitness caused by adapting to the environment with respect to an already adapted individual (specialist). Learning requires time to adapt, defined by the speed of learning $\alpha$. This delay is modeled by reducing the plasticity $\delta$ such that the size of area $A$ is the same.}
  \label{fig7}
\end{figure}

It is assumed that an agent can choose the best skill available \emph{for each resource type} right from the start with no delay, i.e. skill of $a_i+\delta$ for resource type $\alpha_0$ and skill of $a_{i,1}+\delta=1-(a_i-\delta)$ for resource type $\alpha_1$, which maximize the fitness function.
  The speed of learning, also called time lag, is modeled by reducing the value of $\delta$ (cf. Fig. \ref{fig7}).
In practice the value of $\delta$ depends on the ratio between the speed of learning and the season length: a slower learning mechanism reduces the distance to which the value can change, similarly a shorter season reduces the number of experiences an agent can have during a season.

\begin{figure}[H]
\centering
\begin{minipage}[t]{0.45\linewidth}
  \includegraphics[width=\textwidth]{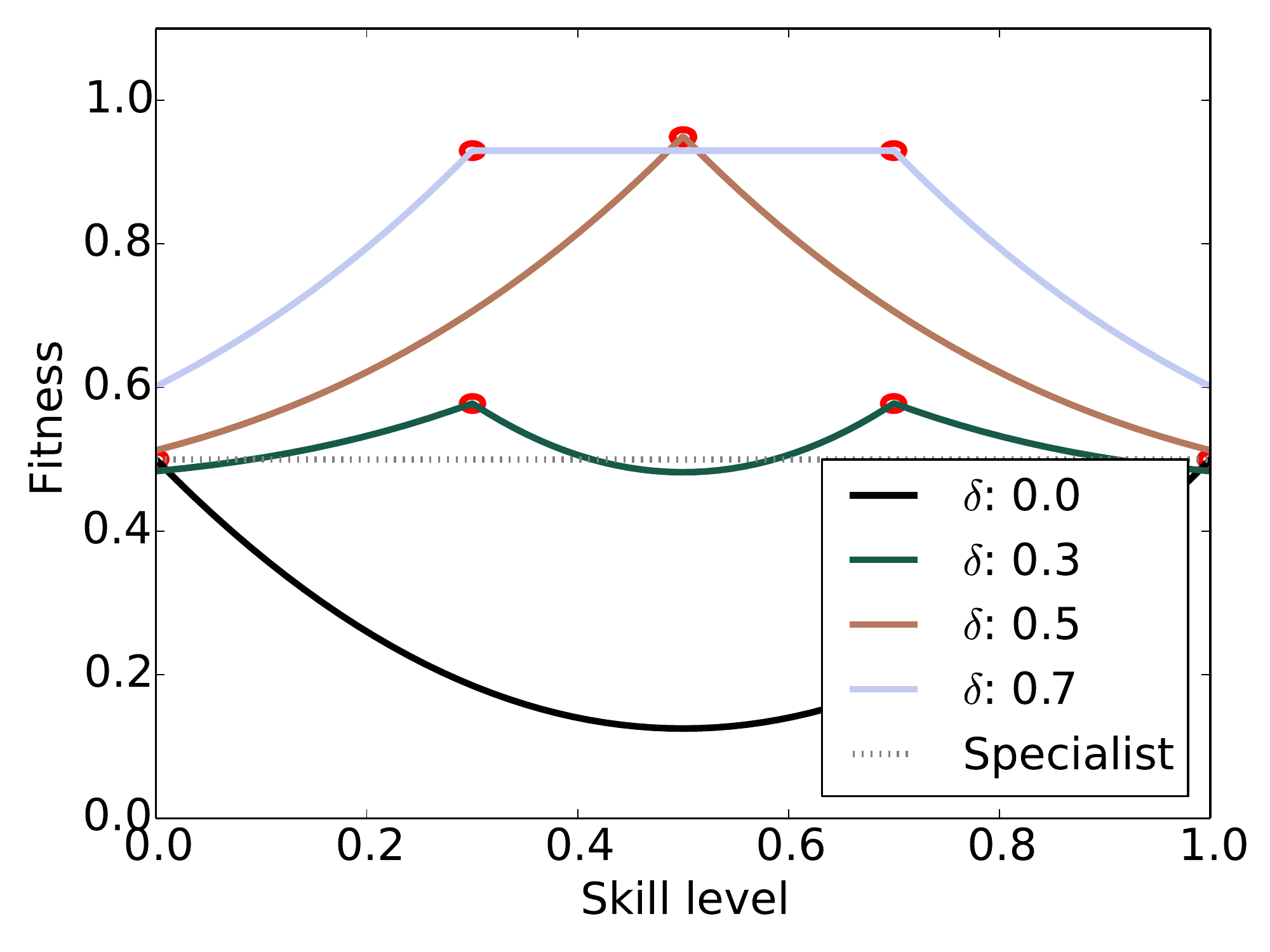}
    \subcaption{Results for $a_0=0.5$.}
\end{minipage}
\begin{minipage}[t]{0.45\linewidth}
  \includegraphics[width=\textwidth]{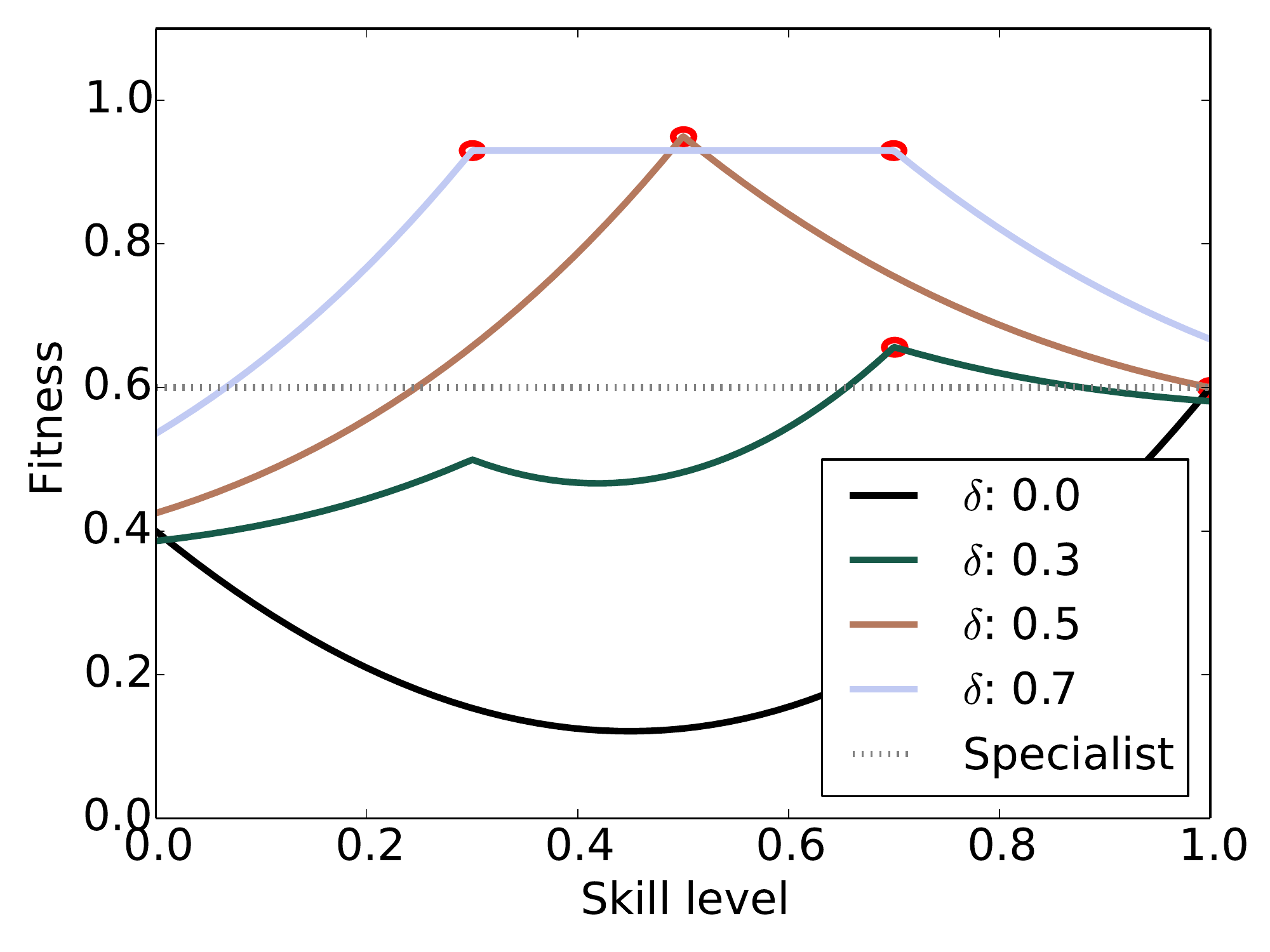}
    \subcaption{Results for $a_0=0.6$.}
\end{minipage}
\caption{{\bf Fitness corresponding to different combinations of aptitude and $\delta$, for $q>1$ and $c>0$} The plot shows the fitness predicted by the mathematical model for given values of aptitude and $\delta$. Left: $a_0=0.5$, right: $a_0=0.6$. The red circles represent the maximum fitness achievable for a given value of $\delta$. Increasing the value of $\delta$ from 0 to 1, the optimal aptitude values start at the extremes (0,1) and move towards the center as $\delta$ increases. The maximum fitness is obtained for $\delta=0.5$, where there is only one maximum for an aptitude of 0.5. For values of $\delta>0.5$ a range of aptitudes, centered in and expanding from 0.5, maximizes the fitness.
  The dotted line corresponds to the fitness of a specialist individual, which becomes lower than the fitness of learning individuals as values of $\delta$ increase.
  Also note that the introduction of learning, i.e. $\delta>0$, changes the aptitude for which fitness is maximized, i.e. the configuration towards which evolution converges.
  \label{fig8}}
\end{figure}

Figure \ref{fig8} shows how different aptitudes compare, in terms of fitness, for varying values of $\delta$.
The red circles represent the globally optimum aptitudes for a given value of $\delta$. If $\delta<0.5$ agents evolve a specialist configuration, as opposed to a generalist configuration if $\delta=0.5$. Note that the configuration with $\delta=0.5$ and aptitude $a_i=0.5$ maximizes the fitness as it allows agents to choose any skill value in the range $[0,1]$, hence allows agents to forage both resource types with certainty.
This condition is verified in practice when the speed of learning is as fast as the frequency of change in the environment, i.e. and agent can adapt its skill to a new environmental state but does so too slowly to remain specialized for a long time before the environment changes again.
  This confirms the existence of the ``Baldwin veering effect'', as any value of $\delta>0$ changes the fitness landscape such that fitness is maximized by a different aptitude, which is then selected.

  For values of $\delta>0.5$, learning makes an increasingly large range of aptitude values equivalent in terms of evolutionary fitness which could allow agents to generalize, but such a configuration would not evolve in reality as the overall fitness is reduced when compared to $\delta=0.5$.
  These results are confirmed also for $c=0$ and $q \leq 1$, see the supplementary material, section \ref{sec:APsimplemodel}.

Concluding, learning agents evolve an intermediate aptitude, i.e. a generalist configuration, only if learning speed is proportionate to the season length such that agents can adapt to both resource types.
This result is general and hold independently of the value of $q$ and resource proportion $\alpha_0$, hence confirms that the Baldwin veering effect depends exclusively on the timescales of learning and environmental change.

\section{Discussion \label{lab:discussion}}

A common finding in the literature about the interactions between reversible plasticity and cyclically-changing environments is that when the frequency of change in the environment is faster than a certain threshold, plastic individuals, who can adapt to changes in the environment after a certain time lag i.e. speed of learning, are more fit than static individuals, who are unable to adapt.
The definition of plasticity varies in the literature: plasticity is modeled as switching between two distinct phenotypes \cite{Padilla1996,Gabriel2005}, as a change in niche breadth \cite{Kassen2002}, or as behavioral adaptation through learning \cite{Floreano1997,Wakano2004}. This work adopts the latter definition.
It is important to note that although the concepts of specialization, i.e. adaptation to only one state, and generalization, i.e. adaptation to more than one state, are consistent across the literature, the concepts of generalist and specialist differ substantially:
In previous work, e.g. \cite{Gabriel2005}, specialist agents are allowed to specialize in two environmental states, hence they are equivalent to a generalist individual that is able to learn, as defined in this paper.

This paper confirms previous results and extends them by introducing the aspect of genetic assimilation: to the best of our knowledge, this work is the first to investigate the effect of reversible plasticity on the evolved genome due to cyclical changes in the environment.
Previous work \cite{Padilla1996,Wakano2004,Kassen2002} investigates the evolution of plasticity by comparing different types of plasticity, in these experiments the genome is fixed and not allowed to evolve.
Other work \cite{Gabriel2005} predicts that non-plastic individuals are better off when able to cope with any environmental configuration, hence they would evolve a genotype that leads to a wide tolerance function. This result conflicts with the finding that static individuals would specialize in either environmental configuration and the population would split in two groups with opposite specialization, we believe this is caused by a modeling assumption that is relaxed in this paper, i.e. each agent can express a different phenotype (tolerance function) for each environmental state.
Other work \cite{Floreano1997,Nolfi1996} looks at the evolution of artificial neural networks and concludes that different levels of plasticity lead to the evolution of different weights.
The main difference with the proposed model is that the environment does not change \cite{Floreano1997} or changes from one generation to the next \cite{Nolfi1996}, hence that model is not able to capture the effect of the frequency of environmental change on evolution.

The aim of this paper is to provide a proof of concept, not modeling realistic entities, hence the model is constrained to only two resources.
Increasing the complexity of the environment, as well as introducing group behavior, is required to model any realistic ecosystem and is left for future work.
The complex interactions between genes that lead to plasticity are simplified to a single gene called aptitude, this assumption has the advantage of letting us develop a simple mathematical model but has the potential disadvantage of being a limitation that conditions the results of the model.
We believe such a binary trade-off is a reasonable simplification as it appears a good model for some simple natural organisms e.g. fish behavior \cite{chapman2009plasticity} and foraging in bacteria \cite{gottschal1979competition}.

Future work will focus on verifying the predictions of the analytical model within the agent-based simulation framework, in particular that it exists a configuration for which a learning population splits in two groups of specialists with aptitude values in $[0,0.5]$ and $[0.5,1]$ respectively, and a configuration in which learning population evolve a uniform distribution of aptitude values.

\section{Conclusions \label{lab:conc}}

Reversible plasticity, e.g. learning, is known to influence the speed at which evolution converges to some optimal configuration.
This work, in contrast, addresses the question of whether or not reversible plasticity in a cyclically-changing environment leads to the evolution of a genetic configuration that differs from what would evolve with a fixed phenotype.
Following previous work, this question is answered by means of an agent-based model of a foraging task, with cyclical variability in the resource distribution. Additionally, this result is confirmed through an analytical model.

Experimental and analytical results show the existence of the Baldwin effect in a cyclical environment and identifies the novel ``Baldwin veering effect'' and the conditions under which it exists.
More specifically an environment that changes cyclically at a speed faster than an individual's lifetime, allows learning agents to evolve a generalist configuration in the same conditions where reactive agents would evolve a specialized configuration.
A mathematical model verifies that the introduction of plasticity in the phenotype changes the fitness landscape in a way such that a generalist strategy becomes the global optimum in the space of genotypes.

These results are relevant for the literature of Evolutionary Biology, as they expand the understanding of how phenotypic plasticity influences evolution and open up a novel dimension for the study of the interaction between learning and evolution.
These results might be relevant in other context which have a cyclical component, for example they might help to understand how technology, e.g. machine learning \cite{pariser2011filter,nguyen2014exploring} which operates at a much faster timescale than human reasoning, might be used to influence opinion formation and polarization by mediating the rate of exposure to different opinions \cite{mas2013differentiation,quattrociocchi2014opinion}.

\section{Methods \label{lab:method}}

\begin{table}[H]
  \centering
  {\footnotesize
  \begin{tabularx}{\columnwidth}{@{}lX@{}}
    Math symbol & Description \\
    \hline
    $\mathcal{A}=\{a_0,...,a_N\}$ & The set of all $N$ agents ever alive in the simulation\\
    $R=\{r_0,...,r_M\}$ & The set of $M$ resource types \\
    $F^t=\{ \phi_{i,r}^t \in E^t : \phi_{i,r}^t >0 \}$ & Set of all cells containing resources\\
    $\phi^t_{i,r} \in \mathbb{N}_{\geq 0}$ & The quantity of resources of type $r$ in cell $i$ at time $t$\\
    $E^t=\{\phi_{i,r}^t : 1 \le i \le m \times m, r  \in R \}$ & The configuration of the environment at time $t$\\
    $T=\{ t \in \mathbb{N}_{0,\leq L} \}$ & The time steps, $t$ of the simulation\\
    $s_a^t \in [0,1] ~:~ a \in \mathcal{A}^t$ & The skill level of agent $a$ at time $t$\\
    $\mathcal{A}^t=\{a\in \mathcal{A} : a $ \mbox{is alive at timestep} $ t\}$ & The population at time $t$ \\
    $f(a,t): \mathcal{A}^t \times T \rightarrow \mathbb{R}$ &  The fitness function\\
    $\epsilon \in \mathbb{R}$ & Energy level increased by successful foraging\\
    $g(a,s_a^t,r):\mathcal{A}^t \times \mathbb{R}_{\geq 0,\leq 1} \times R \rightarrow \{0,1\}$ & The foraging success function of agent $a$ for resource type $r$\\
    $B(a,t):\mathcal{A}^t \times T \rightarrow O$ & The decision function which determines the behavior of agent $a$ at time $t$\\
    $O=\{ o_1,...,o_n\}$ & The set of $n$ possible actions\\
    $P_f(a,t,r):\mathcal{A}^t \times T \times R  \rightarrow [0,1]$ & The probability at time $t$ of agent $a$ to forage resources of type $r$\\
    $P_r(a,t):\mathcal{A}^t \times T \rightarrow [0,1]$ & The probability of reproduction of agent $a$ at time $t$\\
    $C_r$ & The normalization constant of reproduction\\
    $P_d(a,t):\mathcal{A}^t \times T \rightarrow [0,1]$ & The probability of death of agent $a$ at time $t$\\
    $d(a,t): \mathcal{A}^t \times T \rightarrow \mathbb{N}_{0,\leq L}$ &  The age function\\
    $C_d$ & The normalization constant of death\\
    $\mathcal{I}_a^t=\{i \in E^t: i$ \mbox{is visible to} $a$ $\}$ & The perception vector of agent $a$ at time $t$\\
    $H_{a,r}^t = \sum_{t \ge j \in T_{a,r}} g(a,s_a^j,r)$ & The foraging history of agent $a$ and resource type $r$ at time $t$\\
    $T_{a,r}=\{t \in T : a$ \mbox{choses to eat} $r\}$ & The times at which agent $a$ executes a foraging action on a resource of type $r$\\
    $L \in \mathbb{N}_{>0}$ & The simulation length \\
    $l \in \mathbb{N}_{>0}$ & The length of seasons\\
    $H_a^t = \sum_{r \in R} H_{a,r}^t$ & The foraging history of agent $a$ at time $t$\\
    $b:\mathcal{I} \rightarrow \mathbb{R}^n$ & The behavior function which assigns a value to every action\\
    \hline
  \end{tabularx}
  \caption{{\bf Mathematical notation in order of appearance in the text.} Notation for indexes has been slightly abused for the sake of brevity.}
  }
\end{table}

An agent-based simulation framework \cite{epstein99_agent_based_comput_model_gener_social_scien} is developed in which a population of agents $\mathcal{A}$ performs a foraging task \cite{beauchamp00_learn_rules_social_forag,hamblin09_findin_evolut_stabl_learn_rule,2016arXiv160206737B} and is subject to an evolutionary process \cite{perc2013evolutionary}.
  The environment is modeled as a squared grid of size $m \times m$ with continuous boundary conditions in which agents can move.
The environment contains two resource types, i.e. $|R|=2$, whose proportion vary over time \cite{pulliam92_popul_dynam_compl_lands} such that in every ``season'' a specific resource is more abundant than the others.
The number of cells with resources, $|F^t|$, is constant at every point in time: whenever one cell is emptied, a random quantity of resources of the same type spawns at a random location.
The experimental design introduces a trade-off between the foraging success of different resource types, determined by the skill $s_a^t$: agents can either become generalists, i.e. be able to forage both resources with a low probability, or specialize, i.e. be able to forage one resource with a high probability and loose the ability to forage the other.

The energy level of an individual depends on three factors: (i) the availability of resources in the environment at each given time, (ii) the individual skill which determines the probability of successful foraging, and (iii) the individual behavior which determines what actions to execute for a given configuration of the environment.
More formally, the fitness function of an agent $a \in \mathcal{A}$ is defined as $f(a,t)\propto \epsilon \propto (g(a,s_a^t,r), E^t,B(a,t),P_f(a,t,r),s_a^t)$ and is assumed to be directly proportional to the energy gained through foraging, which in turn is proportional to the foraging success.
With a direct relation between skill and probability, i.e. $P_f(a,t)=s_a^t$, the average total intake of an agent is equivalent to the average resource distribution: a specialist agent forages with certainty one type of resources but none of the other, while a generalist agent forages each resource with $50\%$ probability.
Assuming a non-linear relation between skill and foraging probability instead, e.g. $P_f(a,t)=(s_a^t)^q, q > 1$, then a specialization leads to higher fitness than a generalization.

The framework determines the reproduction and death of agent by a genetic operator called \emph{roulette wheel selection with stochastic acceptance} (as in \cite{torney11_signal_evolut_cooper_forag_dynam_envir}), according to which agents reproduce asexually with a probability $P_r(a,t)=f(a,t)/C_r$ proportional to their fitness and die with a probability $P_d(a,t)=d(a,t)/C_d$ proportional to their age.
Upon reproduction, the energy level $\epsilon$ of the parent is split equally between the parent and the offspring and the offspring inherits a randomly-mutated copy of the parent's genetic configuration.

Two types of agents are introduced: reactive agents keep their behavior and skill, a direct expression of their genotype, constant throughout their lifetime, while learning agents adapt according to their experience via reinforcement learning \cite{gruau1993adding,batali96_model_evolut_motiv,redko10_learn_evolut_auton_adapt_agent,nolfi1994phenotypic}.
Learning optimizes the reward associated with successfully foraging a resource of any type.
Different reinforcement learning architectures are evaluated: QLearning \cite{watkins92_q_learn}, reinforcement learning based on a Restricted Boltzman Machine  \cite{hinton06_unsup_discov_nonlin_struc_using_contr_backp}, Deep Reinforcement Learning \cite{mnih2015human} and reinforcement learning based on a single feed forward perceptron (see also the supplementary material, section \ref{sec:learning}).
The behavior of reactive agents is determined by natural selection, while the behavior of learning agents is determined by the interaction between evolution and learning.

The degree of specialization of a population is measured with different metrics:
(I) the distribution of individual aptitudes across the population, according to which a higher frequency of extreme values corresponds to a more specialized population, (II) the individual foraging history, i.e. the frequency of successful foraging actions for a specific resource type, according to which extreme values indicate a specialized diet, (III) standard measures of group behavior that quantify the rate of consumption of resources.
\subsection{Measures}\label{sec:measures}
The degree of specialization of the population is measured by the distribution of aptitudes (I) at each given timestep, normalized by the population size at that timestep.

\[
  M^1(v,t)=|\{a \in \mathcal{A}^t ~:~ s_a^t=v \}|/|\mathcal{A}^t|
\]

The foraging history (II) of the population at value $x$ is measured as the frequency of individuals in the population who, during their lifetime, foraged a specific proportion of type $i$ resources corresponding to $x$.

\[
  M^2(x,r)=| \{a \in \mathcal{A} ~:~ H_{a,r}^L / H_a^L = x\} | / |\mathcal{A}|
\]

Additionally, standard measures of group behavior (III), taken from \cite[Pag.~241]{giraldeau00_social}, are used to quantify the specialization of the population. The measure are defined and explained in \ref{sec:measures}.

While (I) measures the characteristics of the genotype, (II) and (III) measure the behavior of the agents which is determined by the phenotype.
An agent's behavior $B(a,t)=\argmax(b(I_a^t))$ is encoded in its phenotype and associates each perception vector, containing a representation of the surroundings that informs about the presence of resources, to an action.
Note that the phenotype of reactive agents is equal to their genotype.

\section{Acknowledgments}\label{sec:Acknowledgments}
The authors acknowledge support by the European Commission through the ERC Advanced Investigator Grant 'Momentum' (Grant No. 324247).
The authors would also like to thank Leonard Wossnig and Johannes Thiele for their help in developing the simulation framework.

\section*{Appendix}
\appendix
\renewcommand{\thesection}{\Roman{section}}
\section{Model assumptions.}
\label{S3_Appendix}
The analytical model relies on restrictive macroscopic assumptions which enable a straight forward analysis:

\begin{itemize}
\item The fitness of agents is modeled over an abstraction of individual cycles (periods of two seasons that repeat) that removes the time component.
  \begin{itemize}
  \item Available resources are assumed to be constant and equal to the average over a cycle.
  \item Agents do not move, instead they access resources of types $0$ and $1$ with probabilities $a_0$ and $a_1$ respectively.
  \item \emph{Evolution} is not modeled explicitly, instead the evolutionary outcome is inferred from the fitness levels obtained within each cycle.
  \end{itemize}
\item Learning is modeled as skill level plasticity: the parameter $\delta$ determines the range of skill levels an agent can choose at the start of the cycle.
\end{itemize}

\section{Mathematical model sensitivity to different $q$ values}
\label{sec:APsimplemodel}

The results of the analysis presented in the paper are validated for different relationships between the skill level and the foraging success and in absence of plasticity costs.
Results hold hold also if $q=1$ and $q<1$.

\begin{figure}[H]
  \centering
\begin{minipage}[t]{0.45\linewidth}
  \includegraphics[width=\textwidth]{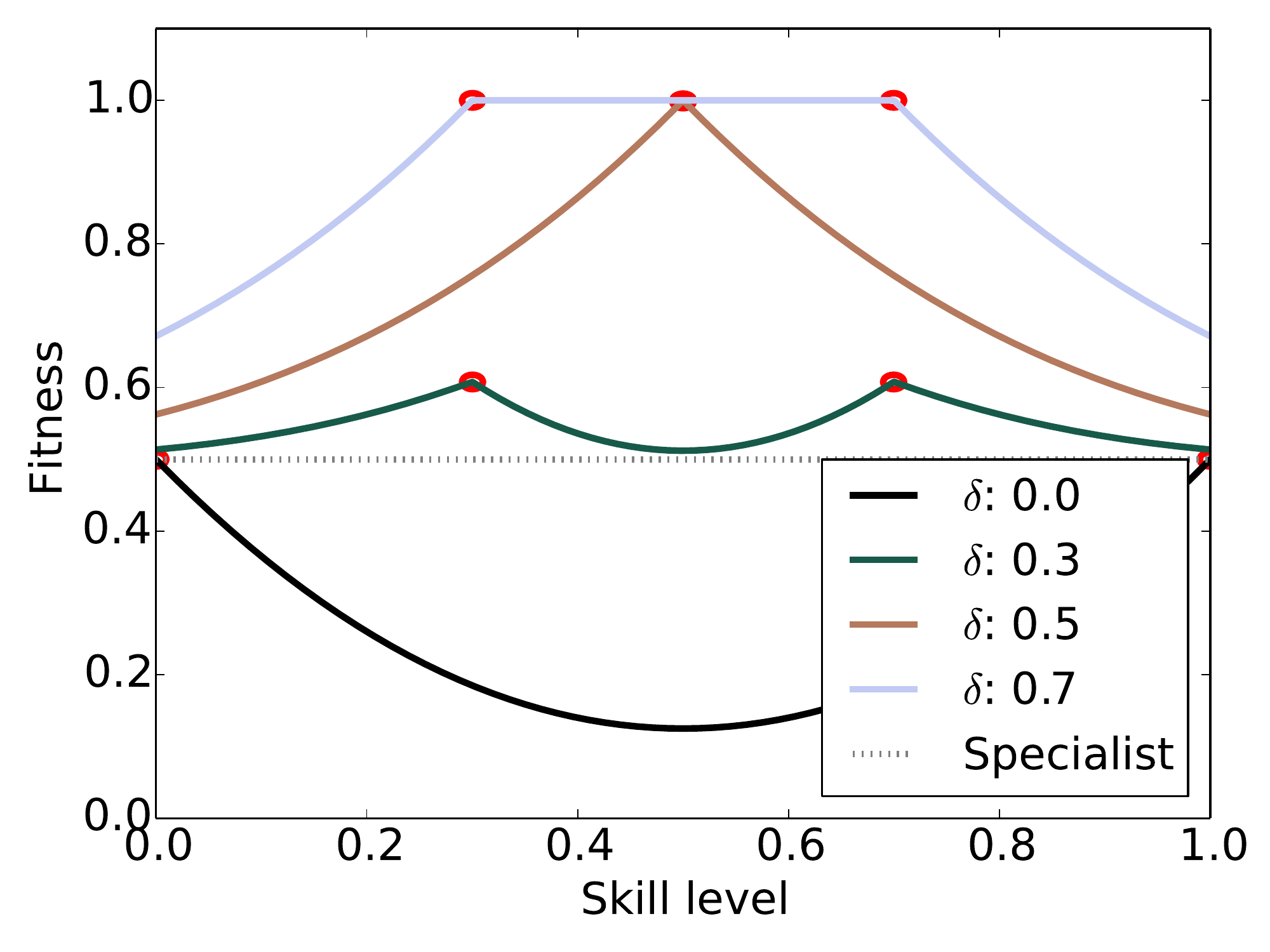}
    \subcaption{Results for $a_0=0.5$.}
\end{minipage}
\begin{minipage}[t]{0.45\linewidth}
  \includegraphics[width=\textwidth]{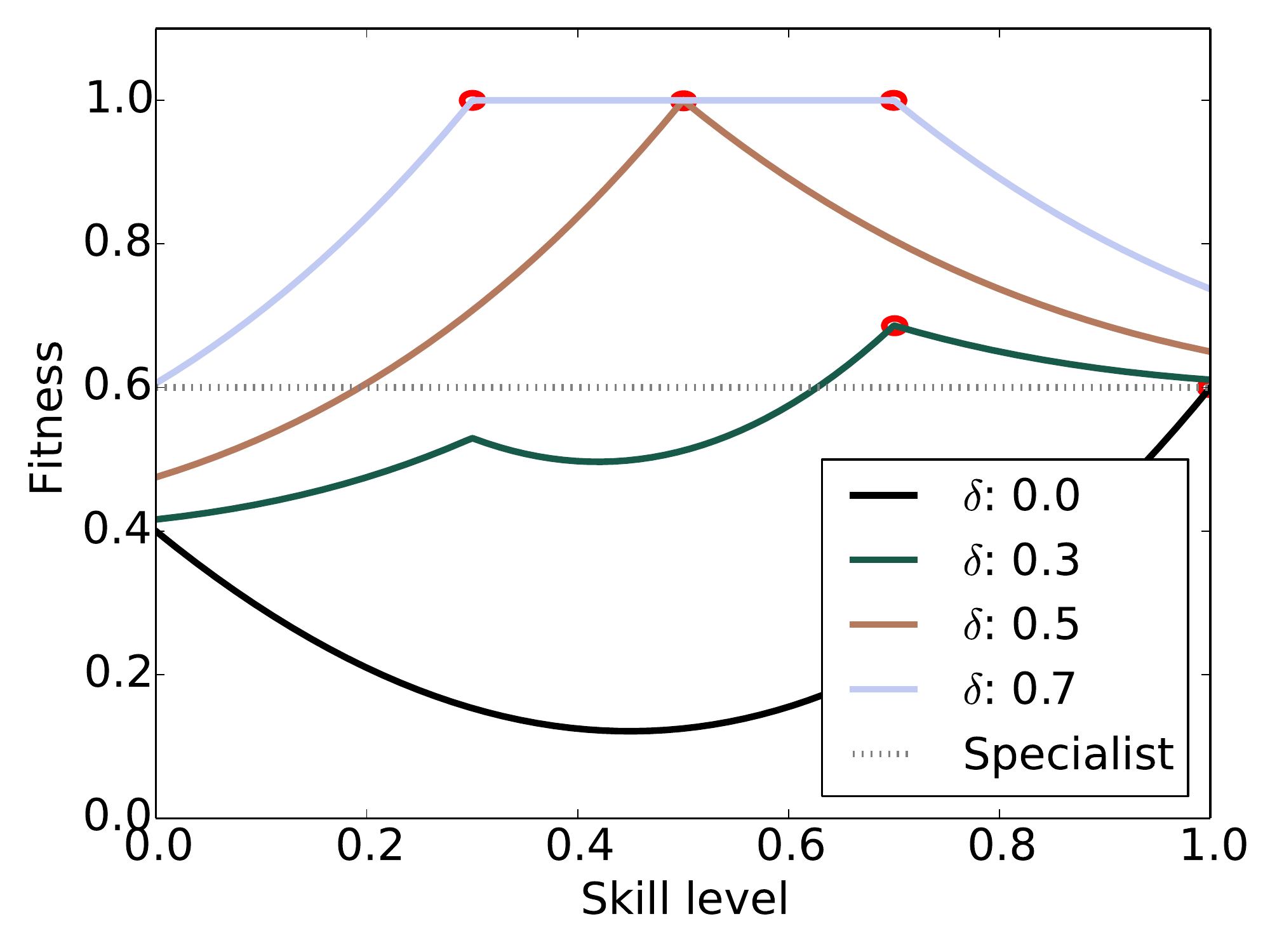}
    \subcaption{Results for $a_0=0.6$.}
  \end{minipage}
  \caption{{\bf Fitness for different combinations of skill level and $\delta$ for $q>1$ and plasticity cost $c=0$}. Note that values of $\delta>0.5$ now maximize the fitness so an evolutionary outcome is possible where a mix of specialists and generalists co-exist.}
  \label{fig9}
\end{figure}

\begin{figure}[H]
  \centering
\begin{minipage}[t]{0.45\linewidth}
  \includegraphics[width=\textwidth]{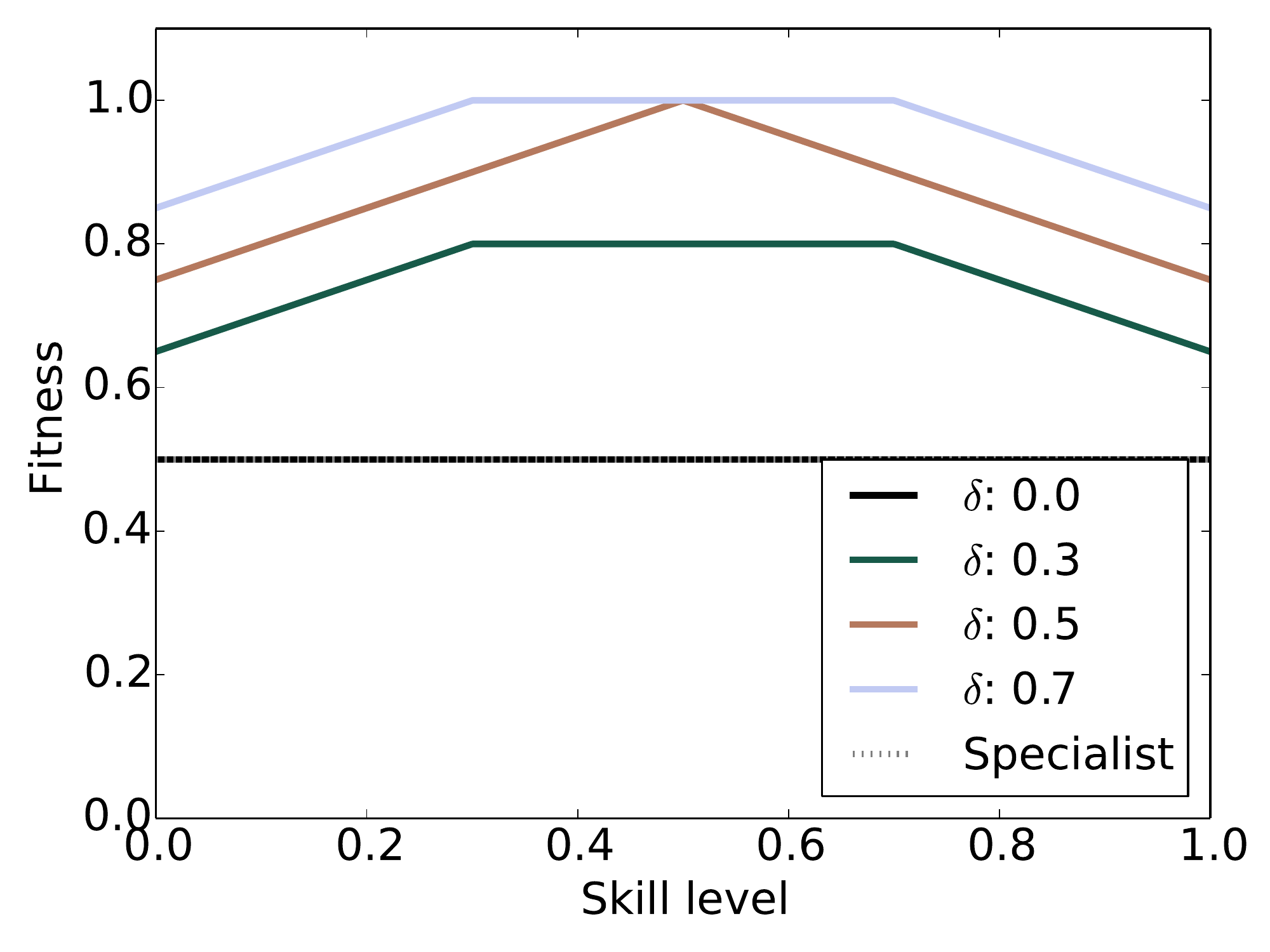}
    \subcaption{Results for $a_0=0.5$.}
\end{minipage}
\begin{minipage}[t]{0.45\linewidth}
  \includegraphics[width=\textwidth]{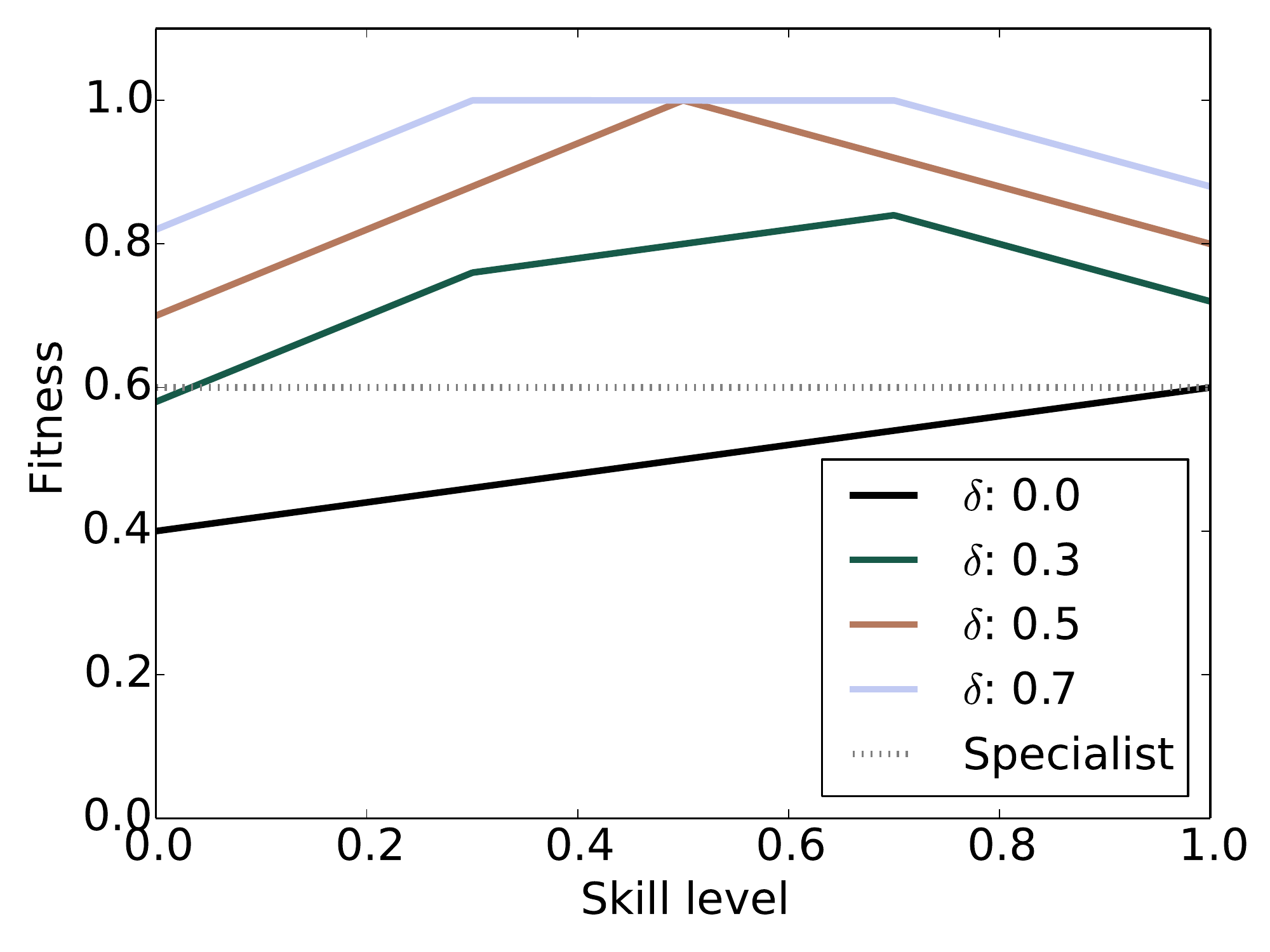}
    \subcaption{Results for $a_0=0.6$.}
\end{minipage}
  \caption{{\bf Fitness for different combinations of skill level and $\delta$ for $q=1$ and plasticity cost $c=0$}. Intermediate skill levels deliver the same fitness as extreme levels, thus a mixed population will evolve. An intermediate skill level of 0.5 is optimal if $\delta=0.5$, while an extreme skill level is optimal for high or low values of}
    \label{fig10}
\end{figure}

\begin{figure}[H]
  \centering
\begin{minipage}[t]{0.45\linewidth}
  \includegraphics[width=\textwidth]{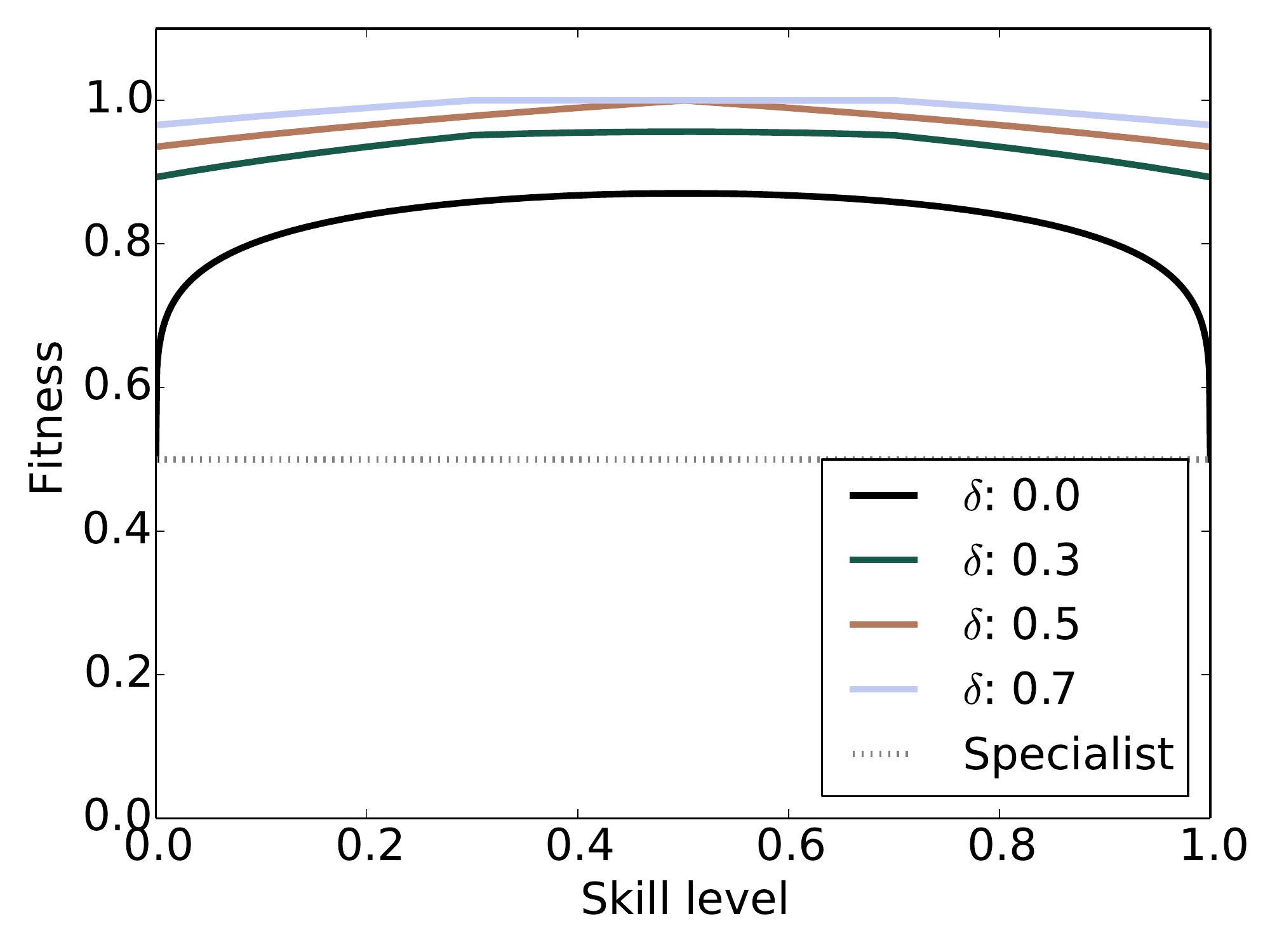}
    \subcaption{Results for $a_0=0.5$.}
\end{minipage}
\begin{minipage}[t]{0.45\linewidth}
  \includegraphics[width=\textwidth]{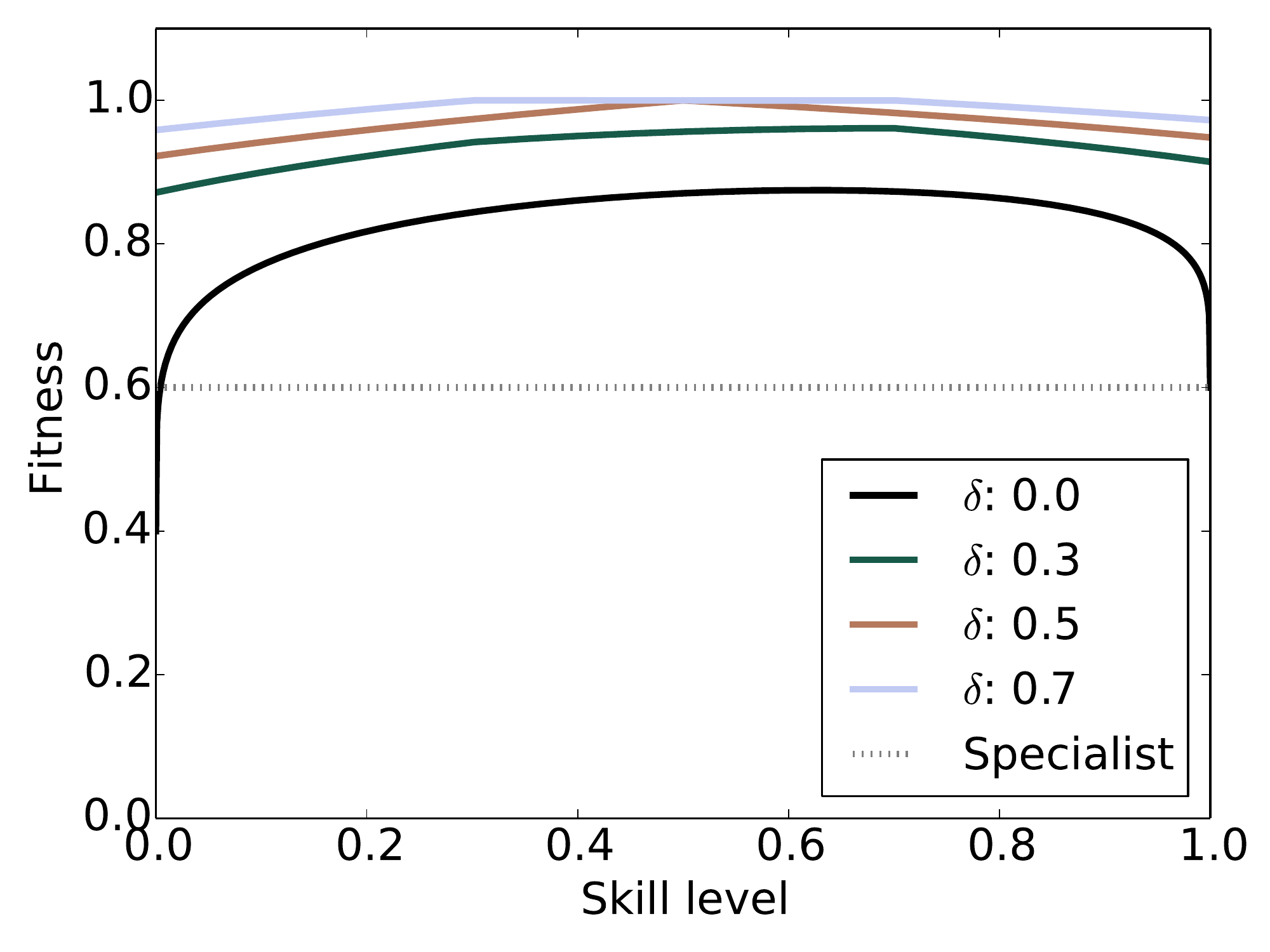}
    \subcaption{Results for $a_0=0.6$.}
\end{minipage}
  \caption{{\bf Fitness for different combinations of skill level and $\delta$ for $q<1$ and plasticity cost $c=0$}. Intermediate skill levels deliver higher fitness than extreme levels, hence specialists have always a lower fitness than generalists. An intermediate a skill level is optimal in any circumstances.}
    \label{fig11}
\end{figure}

\section{Diversity measures for social foraging}
\label{sec:measures}

Assume a group contains $G$ individuals and $S$ discrete resource types.

\begin{itemize}
\item $n_{gs}$ is the number of items of resource $s$ consumed by individual $g$.
\item $n_{g.}=\sum_{s=1}^S n_{gs}$ is the total foraging of individual $g$.
\item $n_{.s}=\sum_{g=1}^G n_{gs}$ is the number of resources of type $s$ foraged by any agent.
\item $n_{..}=\sum_{g=1}^G \sum_{s=1}^S n_{gs}$ is the number of resources of any type consumed by any agent.
\end{itemize}

Each $n_{gs}>0$ defines a sample proportion $p_{gs}$ where $p_{gs}=n_{gs}/n_{..}$, which is used to estimate the total, cross-classified diversity:
$$ h^\prime(g \times s)= - \sum_{g=1}^G \sum_{s=1}^S p_{gs} ln(p_{gs})$$

The following measures of social foraging \cite[Pag.~241]{giraldeau00_social} are based on the concept of diversity \cite{patil1982diversity}:
\begin{itemize}
\item Among-resource diversity: $h^\prime(s)=-\sum_{s=1}^S p_{.s} ln(p_{.s})$
\item Conditional phenotypic diversity within resource $s$: $h^\prime(g|s)=-\sum_{g=1}^G (\frac{p_{gs}}{p_{.s}}) ln(\frac{p_{gs}}{p_{.s}})$
\item Average within-resource diversity: $E[h^\prime(g|s)]=\sum_{s=1}^S p_{.s} h^\prime(g|s)$
\item Among-individual diversity: $h^\prime(g)=-\sum_{g=1}^G p_{g.} ln(p_{g.})$
\item conditional resource-consumption diversity: $h^\prime(s|g)=-\sum_{s=1}^S (\frac{p_{gs}}{p_{g.}}) ln(\frac{p_{gs}}{p_{g.}})$
\item $E[h^\prime(s|g)]=\sum_{g=1}^G p_{g.} h^\prime(s|g)$
\end{itemize}

\begin{figure}[H]
  \begin{framed}
    A generalized diet includes most of all resources types in roughly equal proportions. A specialized diet includes one or a few resource types at high proportions, and very low proportional levels of the remaining resources. The group's diet refers to the pooled resource consumption of all group members.
\begin{itemize}
\item Among-resource diversity $h^\prime(s)=-\sum_{s=1}^S p_{.s} ln(p_{.s})$
  \begin{itemize}
  \item Low: group specializes because individuals have similar specialized diets
  \item High: group generalizes, individuals may generalize or different individuals have different specialized diets.
  \end{itemize}
\item Average within resource diversity $E[h^\prime(g|s)]$.
  \begin{itemize}
  \item Low: different individuals have different specialized diets, so group generalizes; similar effect occurs whenever different individuals consume different total amounts of resources.
  \item High: individuals have similar diets, whether generalized or similarly specialized, group diet may then be generalized or specialized.
  \end{itemize}
\item Among-individual diversity $h^\prime(g)$.
  \begin{itemize}
  \item Low: individuals differ in amount of resources consumed, independently of each individual's specialization or generalization.
  \item High: Individuals consume similar amounts of resources, independently of each individual's specialization or generalization.
  \end{itemize}
\item Average within-individual diversity $E[h^\prime(s|g)]$.
  \begin{itemize}
  \item Low: Individuals specialize independently, group may consequently specialize or generalize.
  \item High: individuals generalize, group consequently generalizes.
  \end{itemize}
\end{itemize}
\end{framed}
\caption{Reproduced from \protect\cite[Pag.~241]{giraldeau00_social}}
\end{figure}

\section{Parameters of the model}
\label{S2_Table}

\begin{table}[H]
  \centering
  \begin{tabularx}{1.0\linewidth}[H]{@{}l|l|X@{}}
    \textbf{Parameter} & \textbf{Value} & \textbf{Description} \\
    \hline
    \textit{Initialization}&&\\
    num-agents & 100 & The size of the initial population. \\
    skill-level & 0.7 & The average skill level of the initial population.\\
    \hline
    \textit{Environment}&&\\
    field-size & 20 & The size of the grid.\\
    max-food & 50 & The maximum resource quantity that a cell can contain. \\
    num-food & 400 & The number of cells containing some food. \\
    food-proportion & 1.0 & The proportion of the 'seasonal' resource with \mbox{respect} to the total amount of resources.\\
    food-energy & 10 & The energy given by a unit of resource. \\
    \hline
    \textit{Agent}&&\\
    max-age & 5000 & Age after which the probability of death is 1.\\
    max-energy & max-age & Age energy after which the probability of \mbox{reproduction} is 1.\\
    fov-radius & 3 & How far agents can perceive.\\
    \hline
    \textit{Simulation}&&\\
    sim-length & 5001 & The length of the simulation.\\
    max-agents & 2000 & The maximum population size, enforced by killing random agents in surplus. \\
    samples   & 50 & The number of independent simulations.\\

  \end{tabularx}
  \caption{Description of the parameters in the model and their value.}
  \label{tab:params}
\end{table}

\section{Learning}
\label{sec:learning}
This section discusses how different learning algorithms behave when faced with a variable environment, in terms of convergence and \emph{adaptation to change}.
Different learning algorithms are compared:
\begin{itemize}
\item PQL: Reinforcement learning using a single layer feed forward perceptron as its network architecture to "store" and query the Q-values
\item RQL: Reinforcement learning using a variation of a Restricted Boltzmann machine \cite{hinton06_unsup_discov_nonlin_struc_using_contr_backp} for the network architecture
\item Q-Learning \cite{watkins92_q_learn}
\item Deep Reinforcement Learning \cite{mnih2015human}: using 3 fully connected layers, $(perception\_size\times perception\_size*5),(perception\_size*5\times number\_of\_actions*5),(perception\_size*5\times number\_of\_actions)$ using gradient descent and action replay with a memory replay of 50 experiences.
\end{itemize}
The results of each learning algorithm are the average of 300 independent simulations, parameters are consistent across simulations.

\begin{figure}[H]
  \centering
  \includegraphics[width=0.9\textwidth]{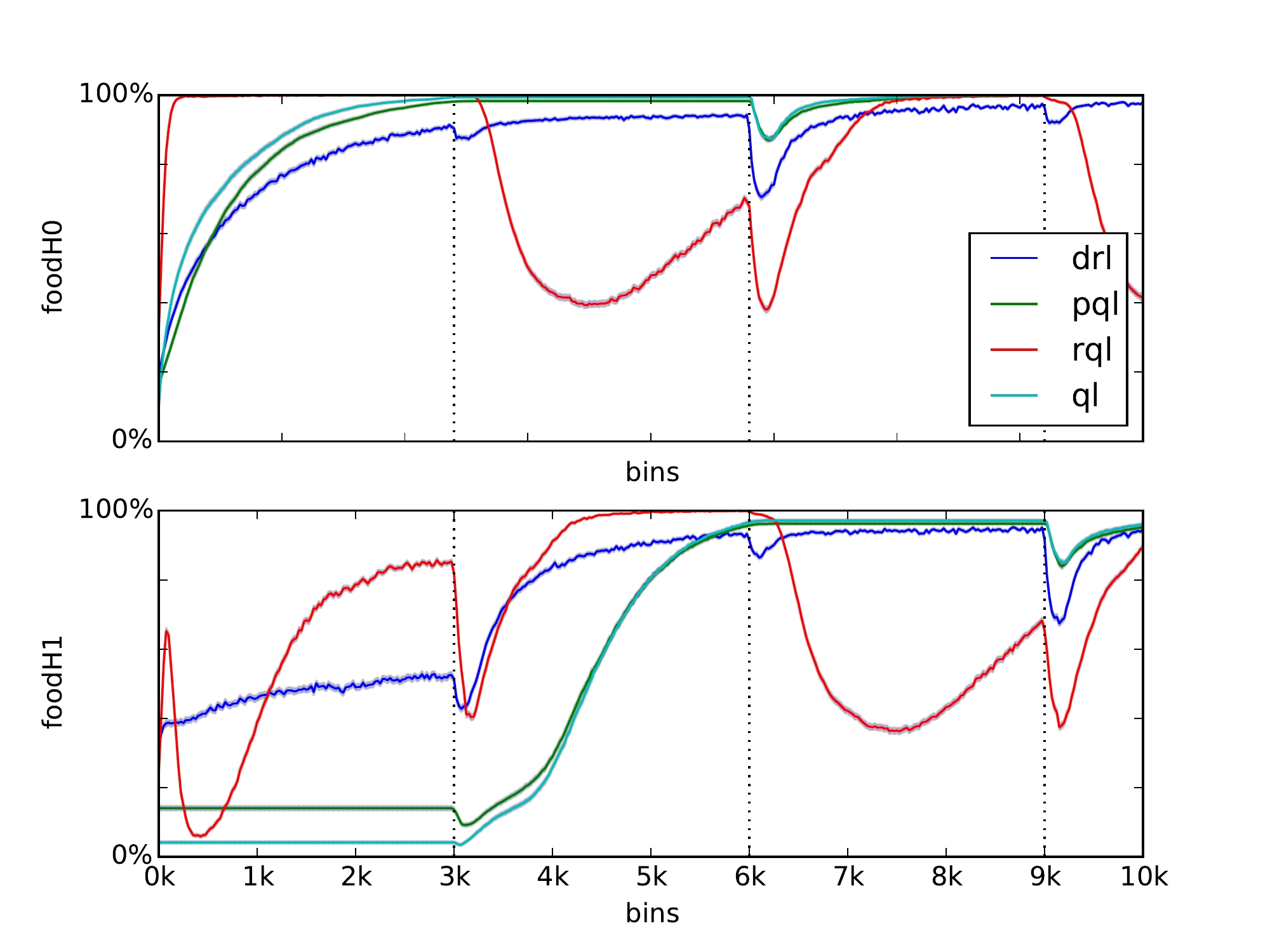}
  \caption{{\bf Comparison of different learning algorithms.} Each graph represent the frequency over time of an agent choosing to forage each resource type whenever the corresponding resource is available.
    A higher value produces a higher fitness, assuming the corresponding resource is available in the environment.
    Each curve is the average of 300 independent simulations. Season length is 3000 and all simulations start in the same season.}
  \label{fig13}
\end{figure}

Results show that different types of learning algorithm have different speeds of convergence (cf. Fig. \ref{fig13}) shows the proportion of agents choosing to eat while a specific type of resource is in their foraging range.
Some learning algorithms adapt faster than others to changes in the environment.

RQL is the fastest to adapt to a change in the environment, and it also shows a stronger tendency to forget the learned behavior in the opposite season.
DRL is the slowest to learn. This is not surprising as deep networks are generally trained with large datasets and used for much more complex tasks.

\section{The Baldwin Veering Effect and the learning algorithm}
\label{S9_Appendix}

In order to analyze the consistency of the results in respect to the type of learning, experiments have been replicated with different learning algorithms.
As the different algorithms produce quantitatively similar results, RQL has been chosen as the learning algorithm for the experiments presented in the paper.

\begin{figure}[H]
  \centering
\begin{minipage}[t]{0.45\linewidth}
  \includegraphics[width=\textwidth]{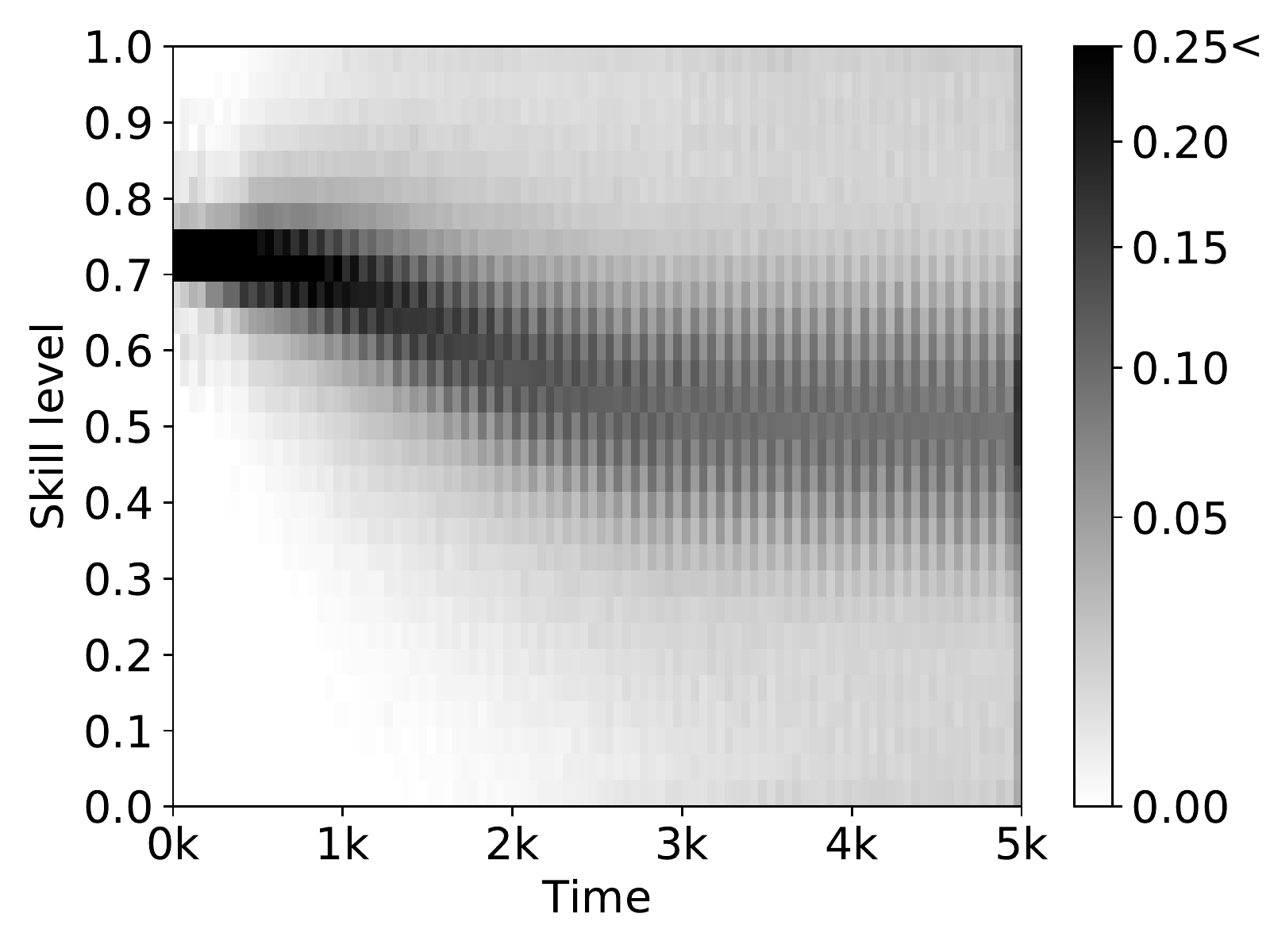}
  \subcaption{PQL.}
\end{minipage}
\begin{minipage}[t]{0.45\linewidth}
  \includegraphics[width=\textwidth]{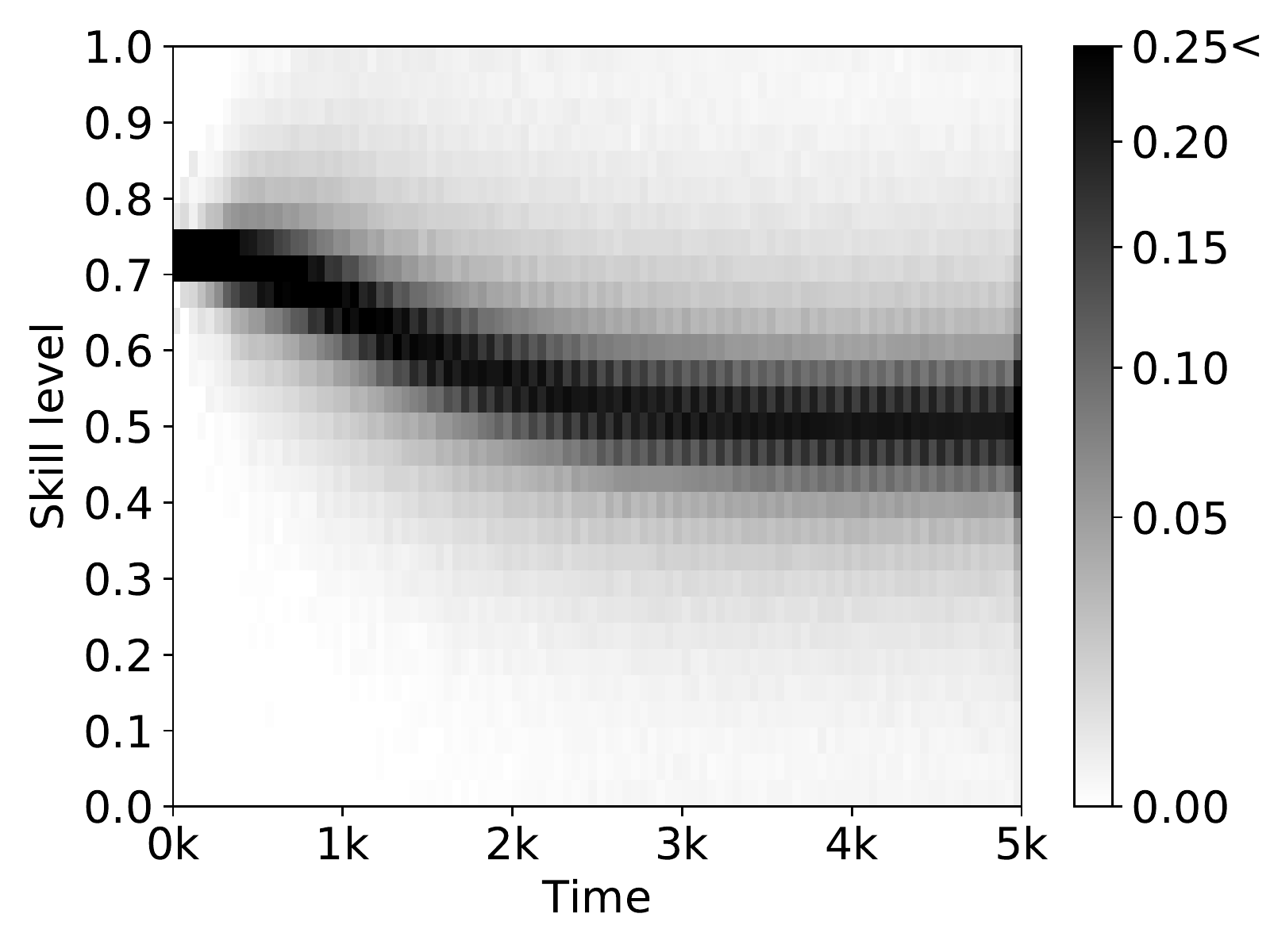}
    \subcaption{QL.}
\end{minipage}
\begin{minipage}[t]{0.45\linewidth}
  \includegraphics[width=\textwidth]{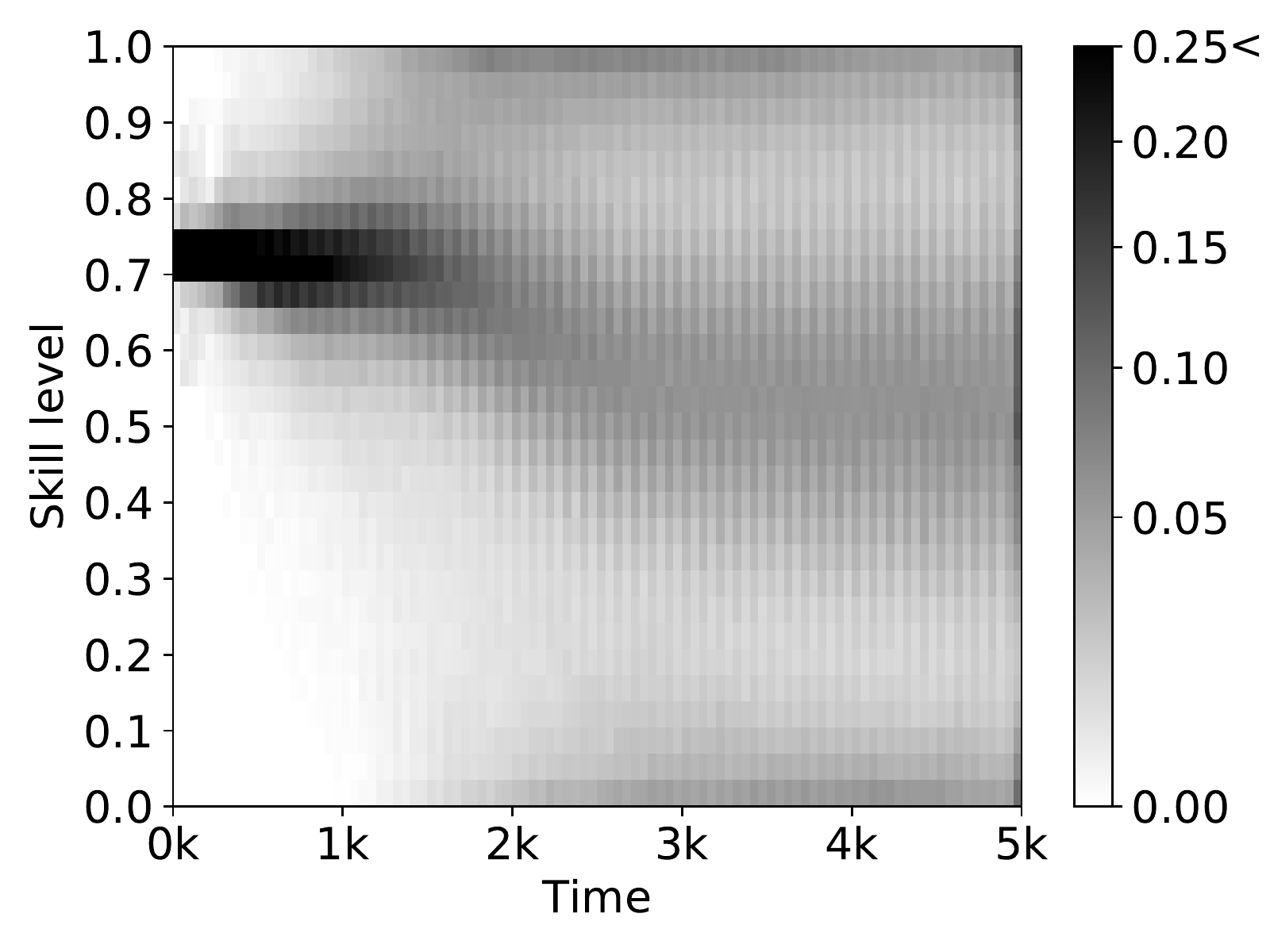}
    \subcaption{RQL.}
\end{minipage}
\begin{minipage}[t]{0.45\linewidth}
  \includegraphics[width=\textwidth]{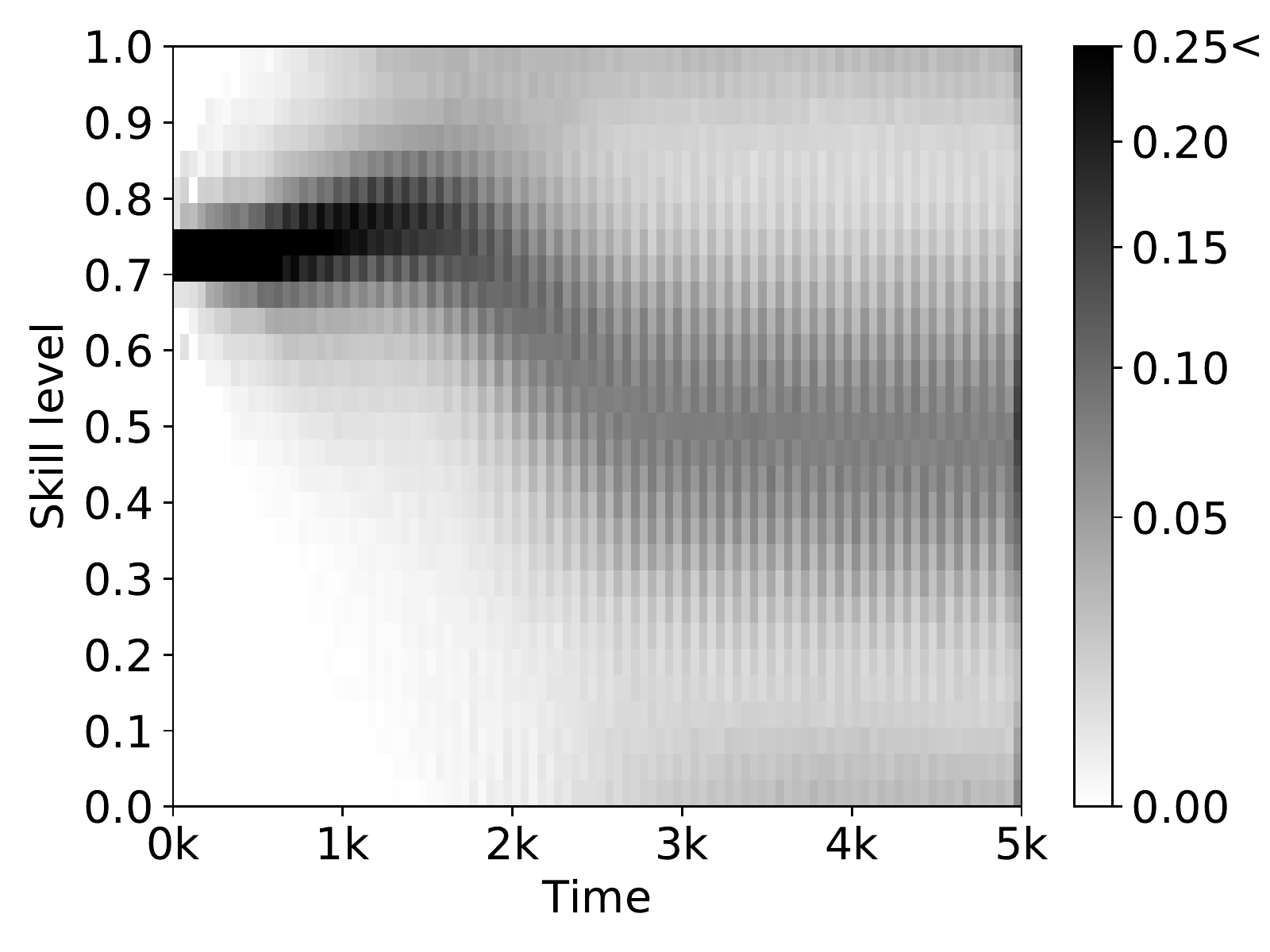}
    \subcaption{DRL.}
\end{minipage}
  \caption{{\bf The genetic configuration evolved with different learning algorithms.} All tested algorithms produce qualitatively similar results.}
  \label{fig14}
\end{figure}

\end{document}